\documentclass[journal]{IEEEtran}
\usepackage[table]{xcolor}
\usepackage{graphicx}
\usepackage{booktabs}
\usepackage{latexsym,bm,amsmath,amssymb}
\usepackage{epsfig}
\usepackage{graphicx}
\usepackage{subfigure}
\usepackage{multirow}
\usepackage{ragged2e}
\usepackage[linesnumbered,ruled,lined]{algorithm2e}
\usepackage{color}
\usepackage{xcolor}
\usepackage{colortbl}
\usepackage{mathrsfs}
\usepackage{array}
\usepackage{enumerate}
\usepackage[misc]{ifsym}
\usepackage{cite}
\usepackage{tikz}
\usepackage{makecell}

\usepackage{cases}
\usepackage{threeparttable}
\usepackage[switch]{lineno}
\usepackage[colorlinks, linkcolor=red]{hyperref}

\usepackage{amsthm}
\SetKw{Continue}{continue}
\SetKw{Boolean}{boolean}
\SetKw{Return}{return}

\DeclareMathSizes{10}{9}{7}{5}

\ifCLASSINFOpdf
\else
\fi
\begin{document}
%
\title{KC-BFPRL: Knowledge-Guided Multi-UAV Collaboration for Grassland Restoration via Bilevel Formerpointer-Based Reinforcement Learning
}

%
\author{\IEEEauthorblockN{Dongbin~Jiao\IEEEauthorrefmark{1}\IEEEauthorrefmark{2}, \IEEEmembership{Member,~IEEE,}
Xianyi~Wang\IEEEauthorrefmark{1}, Yuchen~Yuan\IEEEauthorrefmark{1}, Weibo~Yang\IEEEauthorrefmark{5}, Peng Yang\IEEEauthorrefmark{4}, \IEEEmembership{Senior Member, IEEE}, Peng~Zhao\IEEEauthorrefmark{1},  
Zhanhuan Shang\IEEEauthorrefmark{6}, and Shi Yan\IEEEauthorrefmark{1}, \IEEEmembership{Senior Member, IEEE}}
\thanks{\IEEEauthorblockA{\IEEEauthorrefmark{1}School of Information Science and Engineering, Lanzhou University, Lanzhou, 730000, P. R. China (e-mail: \{jiaodb, wxianyi2025, yuanych2020, zhaopeng, yanshi\}@lzu.edu.cn).}}
\thanks{\IEEEauthorblockA{\IEEEauthorrefmark{2}Key Laboratory of Tourism Information Fusion Processing and Data Ownership Protection, Ministry of Culture and Tourism, Lanzhou University, Lanzhou, 730000, P. R. China.}}
\thanks{\IEEEauthorblockA{\IEEEauthorrefmark{4}Department of Statistics and Data Science, Southern University of Science and Technology, Shenzhen 518055, P. R. China (e-mail: yangp@sustech.edu.cn).}}
\thanks{\IEEEauthorblockA{\IEEEauthorrefmark{5}School of Automobile, Chang'an University, Xi'an, 710064, P. R. China (e-mail: wbyang@chd.edu.cn).}}
\thanks{\IEEEauthorblockA{\IEEEauthorrefmark{6}State Key Laboratory of Grassland Agro-Ecosystem, College of Ecology, Lanzhou University, Lanzhou, 730000, P. R. China (e-mail: shangzhh@lzu.edu.cn).}}
}
\maketitle

\begin{abstract}
Multi-unmanned aerial vehicle (UAV) systems provide scalable service platforms for large-scale environmental tasks, such as grassland ecosystem restoration. However, coordinating fleet operations requires solving the  restoration area maximization problem (RAMP). This non-linear combinatorial optimization challenge is complicated by payload-dependent energy dynamics and heterogeneous ecological degradation. We propose a novel knowledge-guided collaborative bilevel formerpointer reinforcement learning framework (KC-BFPRL) to address this complexity. Using a hierarchical paradigm, KC-BFPRL decomposes RAMP into global task allocation and local restoration planning, with the latter further divided into upper-level trajectory planning and lower-level restoration area allocation. Our specialized architecture pairs featuring a Transformer-based encoder that fuses static environmental features with dynamic UAV states, and a Pointer Network decoder trained via a robust actor-critic framework. By embedding ecological priority rules and heuristic logic, KC-BFPRL achieves a structured warm-start, solving the RL cold-start problem while ensuring strict constraint satisfaction.  Extensive experiments demonstrate that KC-BFPRL consistently outperforms state-of-the-art baselines, achieving superior objective values and efficiency. It maintains a $0.00\%$ optimality gap in the most complex scenarios U8-R160 and operates nearly three times faster than MAPDP, validating its robustness, scalability, and real-time applicability for large-scale automated ecological restoration.
\end{abstract}

\IEEEpeerreviewmaketitle
\section{Introduction}
Grassland ecosystems are vital for global ecological stability, yet accelerating environmental degradation demands scalable and efficient restoration interventions~\cite{sommer2023grassland,waring2024grand}. While traditional manual methods are prohibitively labor-intensive, the integration of unmanned aerial vehicles (UAVs) has introduced a transformative paradigm in ecological engineering by offering high operational flexibility and low deployment costs. Despite these advantages, current UAV frameworks are predominantly designed for passive monitoring. Transitioning to active restoration, such as aerial seeding, introduces profound operational complexities. Unlike lightweight monitoring missions, carrying heavy seed payloads drastically depletes battery reserves and constrains aerodynamic maneuverability. Consequently, the severe energy and payload limitations of a single UAV render it insufficient for executing large-scale restoration across degraded terrains
\cite{JIAO2024108084}.

To overcome this bottleneck, a strategic shift toward multi-UAV collaborative systems is imperative~\cite{rizk2019cooperative,yang2025joint}. While fleet coordination enables parallel execution, orchestrating such a system introduces immense computational complexity, formulated here as the restoration area maximization problem (RAMP). Distinct from standard vehicle routing problem (VRP) formulations, RAMP represents a highly coupled, non-linear combinatorial optimization challenge characterized by heterogeneous degradation levels and stringent resource constraints \cite{JIAO2024108084}. The system dynamics exhibit strong nonlinearity and intricate coupling among decision variables: different regions require distinct seeding densities, while a UAV's energy expenditure is non-linear, decreasing progressively with payload discharge. As a result, RAMP entails a tightly coupled joint optimization of task allocation and path planning, with the primary objective of maximizing the total restored areas before battery depletion.

Existing approaches to solving such complex collaborative problems generally fall into two categories: heuristic optimization and deterministic methods \cite{lahrichi2015integrative}. Traditional meta-heuristics (e.g., Genetic Algorithms (GA), Particle Swarm Optimization (PSO)) are computationally lightweight but demand substantial domain expertise, lack cross-scale generalization, and frequently stagnate in inferior local optima~\cite{verdu2025scaling}. Conversely, deterministic methods struggle with the high dimensionality and dynamic state spaces of multi-agent environments, rapidly becoming computationally intractable as fleet and network sizes expand \cite{huo2022two}.

To address the longstanding trade-off between solution quality and computational efficiency, this paper proposes a novel framework based on deep reinforcement learning (DRL). DRL is particularly suited to this problem due to its capacity to handle high-dimensional state spaces and learn nonlinear decision policies directly from interaction data \cite{liang2025enhancing}. Building on this capability, we introduce  the knowledge-guided collaborative bilevel formerpointer reinforcement learning framework, termed KC-BFPRL. To effectively fuse static environmental features with dynamic UAV states, KC-BFPRL incorporates a specialized architecture featuring a Transformer-based encoder and a Pointer Network decoder trained via a robust actor-critic mechanism. Departing from inefficient, undirected exploration, KC-BFPRL utilizes ecological priority rules and heuristic scheduling logic as an inductive scaffold \cite{zhang2024knowledge}. This structured domain knowledge explicitly addresses the cold-start problem, significantly enhancing sample efficiency and policy feasibility \cite{roy2020promoting}.

To implement this vision, the grassland restoration process is modeled as a hierarchical Markov decision process (MDP), in which each UAV as an intelligent agent. This design decomposes the complex RAMP into global task allocation and local restoration planning. Specifically, the local planning subproblem is further decoupled into two tightly interdependent levels: upper-level trajectory planning, where UAVs optimize visitation order to minimize energy waste; and lower-level restoration area allocation, which dynamically determines the optimal seeding quantity at each node to balance ecological impact against residual payload and energy reserves. By integrating these levels, KC-BFPRL provides a principled warm-start mechanism that enables UAVs to efficiently acquire complex cooperative behaviors while simultaneously optimizing individual energy states, ultimately achieving enhanced restoration coverage with rapid, real-time inference capability.

The main contributions are summarized as follows: (1) We mathematically model large-scale multi-UAV grassland restoration as RAMP, a complex, non-linear optimization problem coupling routing, task allocation, and payload-dependent energy dynamics. (2) We propose KC-BFPRL, a knowledge-guided bilevel reinforcement learning framework that integrates ecological priorities and heuristic logic into a hierarchical decision-as-a-service architecture. (3) Experiments on 10800 instances demonstrate that The KC-BFPRL 
shows KC-BFPRL outperforms baselines, maintaining a $0.00\%$ optimality gap in complex cases and achieving nearly three times faster inference than MAPDP with robust generalization.

The remainder of this paper is organized as follows. Section \ref{Related-Work} reviews the related literature. Section \ref{System-Model} details the system model and problem formulation. Section \ref{Methodology} introduces the proposed KC-BFPRL framework. Section \ref{Experimental} presents the experimental evaluation, and Section \ref{Conclusion} concludes this work.

\section{Related Work} \label{Related-Work}
This section reviews the methodological shift from traditional heuristics to knowledge-guided, data-driven paradigms for solving RAMP. By examining the intersection of precision agriculture, combinatorial optimization, and MARL, identifying critical research gaps that motivate our KC-BFPRL framework.

\subsection{UAV-based Ecological Restoration and Agriculture.}
The application of UAVs in precision agriculture has evolved significantly over the past decade. Initially, UAVs served primarily as mobile platforms for remote sensing and environmental monitoring \cite{anderson2013lightweight,osco2021review,Rossello2022}. early studies predominantly leveraged hyperspectral imaging to classify vegetation health and assess degradation levels \cite{xu2019responses,nepi2025ai}.

However, recent advances in payload capacity have transformed these platforms from passive observers to active actuators capable of complex tasks like aerial seeding \cite{mukhamediev2023coverage}. Unlike traditional agricultural spraying that assumes uniform coverage \cite{kim2019unmanned,radoglou2020compilation}, grassland restoration involves highly heterogeneous environments where seeding requirements vary strictly by local degradation levels \cite{JIAO2024108084}. Most existing studies focus on simple coverage metrics and fail to account for the complex coupling between a UAV's limited payload, energy constraints, and the spatially varying urgency of restoration demands.

\subsection{Multi-UAV Task Allocation and Path Planning.}
To address the challenges of routing and task allocation for multi-UAV system, early studies rely heavily on mathematical programming and heuristic algorithms. Exact methods, such as mixed-integer linear programming (MILP), can yield global optimal solutions but are constrained by NP-hardness, making them computationally intractable for large-scale scenarios involving dozens of tasks \cite{jones2023path}.

As a result, heuristic and meta-heuristic approaches have emerged as the prevailing solution paradigm. Algorithms including GA, PSO, and ant colony optimization (ACO) have been extensively applied to the VRP and its variants \cite{xie2022multiregional,chen2022coverage,zhang2025dynamic,xu2025evolving}. A representative example is CHAPBILM \cite{JIAO2024108084}, a coupled heuristic framework designed explicitly for bi-level optimization in grassland restoration. Although these heuristic approaches provide stable and interpretable solutions, they incur substantial computational overhead. As the scale of degraded regions expands, their iterative search processes exhibit near-exponential growth in execution time. Furthermore, these methods typically require complete re-optimization from scratch whenever environmental parameters shift. This inherent inflexibility precludes their deployment in time-critical, large-scale scenarios that demand rapid, real-time decision-making and dynamic replanning.

\subsection{DRL for Multi-Agent Collaboration.}
To reduce the computational bottlenecks  of heuristic optimization and deterministic methods, recent research has increasingly shifted toward DRL. Within this paradigm, the joint problem of routing and allocation is formulated as a neural combinatorial optimization (NCO) challenge  \cite{wu2024neural}. Pioneering works, such as Pointer Networks \cite{vinyals2015pointer} and the attention model \cite{kool2018attention}, have demonstrated that neural architectures can learn to construct near-optimal solutions for the Traveling Salesman Problem (TSP) and VRP almost instantaneously following offline training \cite{jiao2026od}.

In the multi-UAV domain, MARL frameworks, such as MADDPG and QMIX, have been adopted to enable decentralized cooperation among agents \cite{rashid2020monotonic}. Recent advances include methods like MAPDP, which exploits context embeddings to decompose large-scale problems \cite{zong2022mapdp}, and CAMP, which leverages attention mechanisms to facilitate inter-agent communication \cite{hua2025camp}.  However, despite their rapid inference speeds, these purely learning-based approaches face profound challenges in strictly constrained restoration scenarios. Lacking prior domain knowledge, agents depend entirely on undirected exploration, resulting in slow convergence and a pronounced cold-start problem. Moreover, their centralized training paradigms often collapse as the number of agents and nodes increases, a fundamental limitation known as the curse of dimensionality. Most importantly, the inherent ``black box" nature of pure DPL makes it difficult to rigorously enforce hard operational constraints, such as energy limits and payload capacities, frequently resulting in infeasible restoration planning.

\subsection{Knowledge-Guided and Hybrid Learning Approaches.}
The inherent limitations of purely heuristic and strictly learning-based approaches have catalyzed a growing interest in knowledge-guided paradigms. These methods aim to embed domain-specific rules or expert priors directly into the learning process to guide state-space exploration and ensure solution feasibility \cite{gao2020knowledge,shahbazian2025knowledge}. Existing hybrid approaches typically depend on heuristics to generate demonstration data for imitation learning or utilize RL to select high-level heuristic operators \cite{bui2023imitation}. However, these existing approaches rarely address the tightly coupled, non-linear constraints characteristic of ecological restoration within a unified, end-to-end architecture.

To bridge these gaps, this paper proposes KC-BFPRL, a knowledge-guided collaborative bilevel formerpointer RL framework for large-scale grassland restoration. Unlike pure MARL methods (e.g., MAPDP), KC-BFPRL explicitly integrates ecological priority and heuristic scheduling logic into the policy learning process, providing a structured warm-start that effectively mitigates the cold-start problem. In contrast to traditional heuristic approaches (e.g., CHAPBILM), KC-BFPRL shifts the computational burden to the offline training phase, enabling rapid real-time inference during deployment. By synergistically combining domain-structured optimality with the adaptability and scalability of DRL, KC-BFPRL offers a robust and efficient solution for constrained, large-scale grassland restoration planning.
\section{System Model and Problem Formulation} \label{System-Model}
This section details the multi-UAV collaborative grassland restoration model, UAV energy consumption dynamics, and the RAMP formulation.
\subsection{Multi-UAV Collaborative Grassland Restoration Model}
\begin{figure}[htb]
\begin{center}
$\begin{array}{l}
\includegraphics[width=2.5in]{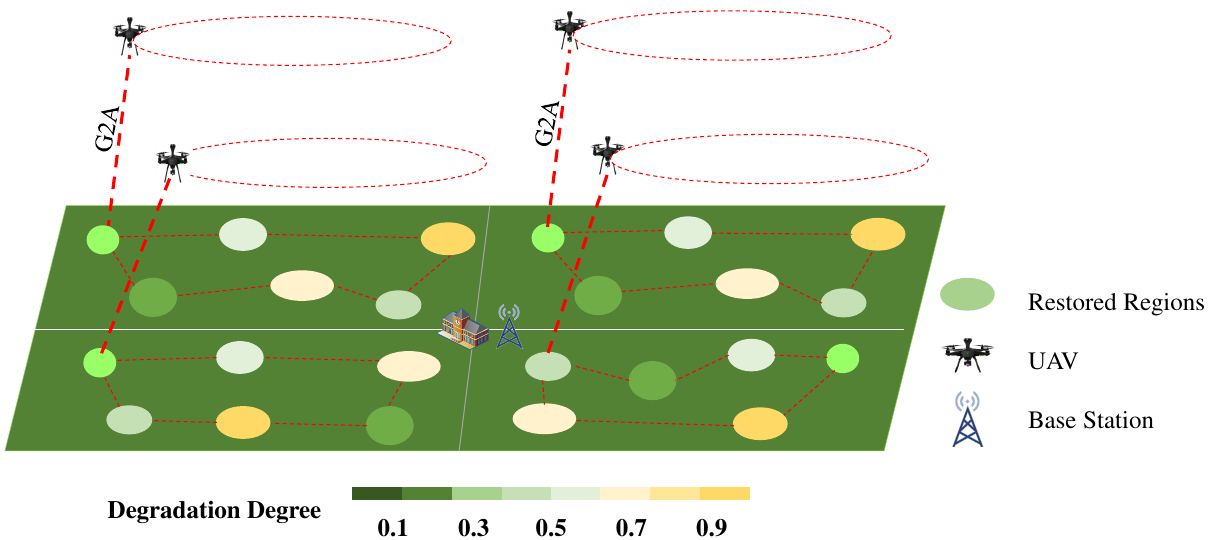}\\
 \end{array}$
\end{center}
\vspace{-0.15in}
\caption{An example of multi-restored regions by multi-UAV collaborative.} \label{multi-UAV-model}
\end{figure}

As depicted in Fig.~\ref{multi-UAV-model}, we considers a scenario where a set $U$ of homogeneous UAVs equipped with BeiDou Navigation Satellite System (BDS) modules and seed dispensers. These UAVs fly at a fixed altitude with a constant flight speed. The mission is coordinated from a central Base Station (BS), which handles scheduling, maintenance, and data processing.

The collaborative restoration process is modeled as a complete directed weighted graph $G=(V,A)$. The vertex set $V=\{v_0,v_1,\ldots,v_N\}$ includes the BS $v_0$ (the depot) and the degraded regions $V_a = V \setminus \{v_0\}$. The arc set $A=\{a_{ij}=(v_i,v_j)|v_i,v_j \in V,i\neq j\}$ represents the flight paths with a Euclidean distance $d_{ij}$.

Each region $v_i$ has a degradation severity level $l_i \in (0,1)$, where a higher score indicates more severe degradation (represented by lighter colors in Fig.~\ref{multi-UAV-model}). Following international ecological restoration standards~\cite{gann2019international}, regions with $l_i < 0.3$ can self-recovery, while those with $l_i > 0.8$ are beyond effective UAV intervention. Consequently, we focuses on the critical interval $l_i \in [0.3,0.8]$, where UAV seeding significantly accelerates recovery and reduces costs~\cite{JIAO2024108084}. Spatially, each region $v_i$ is discretized  into $c_i$ unit circles. A UAV hovers over a circle to sow a seed quantity determined by $l_i$. All UAVs depart from the BS with maximum energy $E_{max}$ and seed payload $Q$, and must complete their tasks and return to the BS before energy depletion.
\subsection{UAV Energy Consumption Model}\label{enegymodel}
Following~\cite{JIAO2024108084}, a UAV's mission energy consumption comprises seeding $E^s$, aerial photography $E^{ap}$, and flight $E^f$ components.

As established in ~\cite{dorling2017vehicle}, flight energy at a constant altitude and speed is directly proportional to the total payload weight.
\subsubsection{Energy Consumption for Seeding}
The total energy required to dispense seeds across all restored regions is formulated as:
\begin{equation} \label{energy-comsumption-seeding}
\begin{small}
E^s=\sum^N_{i=1}\sum^N_{j\neq i}\sigma_i e_i x_{ij},
\end{small}
\end{equation}
where $\sigma_i$ is the number of restored unit circles in region $v_i$, and $x_{ij}$ indicates whether a UAV travels from $v_i$ to $v_j$. The unit seeding energy is $e_i=\eta q_i$, where $\eta > 0$ is a coefficient and
$q_i=(1+l_i)^\gamma$ is the seed weight, with
$\gamma$ being an environmental-specific grassland parameter. 
\subsubsection{Energy Consumption for Aerial Photography}
The total energy consumed by the onboard hyperspectral camera to acquire data at all restored regions is:
\begin{equation} \label{energy-comsumption-seeding}
\begin{small}
E^{ap}=e^{ap}\sum^N_{i=1}\sum^N_{j\neq i}\sigma_i x_{ij},
\end{small}
\end{equation}
where $e^{ap}$ is the data acquisition energy per unit circle.
\subsubsection{Energy Consumption for Flight}
The flight energy depends on travel distance and varying payload weight:
\begin{equation} \label{energy-comsumption-seeding}
\begin{small}
E^f=\sum^N_{i=1}\sum^N_{j\neq i}e^f_{ij}d_{ij} x_{ij},
\end{small}
\end{equation}
where $e^f_{ij}$ is the energy consumption rate per unit distance along edge $(i,j)$, defined as \cite{dorling2017vehicle}:
\begin{equation} \label{energy-consumption-per-unit-distance}
\begin{scriptsize}
e^f_{ij} = P(\bar{q}_{ij}) = (M + \bar{q}_{ij})^{\frac{3}{2}}\sqrt{\frac{g^3}{2 \rho \varsigma h}},
\end{scriptsize}
\end{equation}
where $\bar{q}_{ij}$ is the seed payload from $v_i$ to $v_j$, and $M=W+m$  is the UAV's tare weight~(frame weight $W$ and battery weight $m$). The constants $g$, $\rho$, $\varsigma$, and $h$ denote gravitational acceleration, air density, rotor disc area, and the number of rotors, respectively. 

\subsection{Restoration Area Maximization Model}
While collaborative multi-UAV systems significantly enhance restoration coverage and efficiency through coordinated task allocation, individual UAV remains confronted with stringent individual constraints of limited flight endurance and payload capacity. These persistent energy limitations preclude the complete restoration of all degraded sites in a single mission. Consequently, it is  necessary to develop resource-aware optimization strategies that prioritize ecologically critical regions to maximize the restoration impact within available energy budgets. To quantify this impact, we define the optimization objective function $C$ as follows.
\begin{equation} \label{objectivefunction}
\begin{small}
C = \sum_{i=1}^{N} \left[1 + (l_i - 0.3)\right] \sigma_i,
\end{small}
\end{equation}
where the term  $l_i-0.3$ serves as an ecological weight factor, explicitly prioritizing regions with higher degradation severity levels $l_i$ to maximize the environmental benefit of the intervention.

\subsection{Mathematical Model}\label{methmodel}
The objective of the multi-UAV collaborative scheduling is to maximize the weighted sum of restored areas while strictly adhering to energy and payload constraints. The problem is formulated as a mixed-integer nonlinear programming (MINLP) model:
\begin{subequations}\label{GRP}
\begin{align}
\max_{\begin{subarray}{c}
x_{uij}\\[-1pt]
\sigma_{ui}
\end{subarray}}
&\sum_{u\in  U}\sum_{i=1}^N\sum_{\substack{j\neq i}}^N
(1+l_i-0.3)\,\sigma_{ui}\,x_{uij}
\label{prob:obj}\\[-2pt]
\text{s.t. }\;
& \sum_{i=1}^{N}\sum_{\substack{j\neq i}}^{N} (\sigma_{ui} e_{ui} + e^{ap}\sigma_{ui}) x_{uij} \nonumber \\
& + \sum_{i=0}^{N}\sum_{\substack{j\neq i}}^{N} e_{uij}^f d_{ij} x_{uij}
\le E_{\max}, \quad \forall u\in U, \label{energy-constraint} \\[1mm]
&\sum_{u\in U}\sum_{\substack{i=1\\ i\neq j}}^N
\sigma_{ui}q_{ui}x_{uij}\le Q,\;\forall j\in  V_a
\label{total-load}\\[-2pt]
&\sum_{\substack{j=0\\ j\neq i}}^N\bar q_{uji}
-\sum_{\substack{j=0\\ j\neq i}}^N\bar q_{uij}
=\sigma_{ui}q_{ui},\;
\forall i\in V_a,\forall u \in U
\label{carry-load}\\[-2pt]
&\bar q_{uij}\le Qx_{uij},\;\forall(i,j)\in  A,\forall u \in U
\label{load}\\[-2pt]
&\sum_{u\in U}\sum_{\substack{j=0\\ j\neq i}}^N x_{uji}
=\sum_{u\in  U}\sum_{\substack{j=0\\ j\neq i}}^N x_{uij} = 1,\;\forall i\in V_a
\label{enter-point}\\[-2pt]
&\sum_{j=1}^N x_{u0j} = \sum_{j=1}^N x_{uj0}=1,\;\forall u \in U
\label{start-point}\\[-2pt]
&x_{uij}\in\{0,1\},\;\forall(i,j),\forall u \in U
\label{binary-variable}\\[-2pt]
& 1 \le \sigma_{ui} \le c_{ui}, \quad \sigma_{ui} \in \mathbb{N}^+, \forall i\in V_a,\ \forall u\in U
\label{area-constraint}\\[-2pt]
&\bar q_{uij}\ge0,\;\forall(i,j),\forall u \in U.
\label{flight-weight}
\end{align}
\end{subequations}
Here, the binary decision variable $x_{uij}$ equals 1 if UAV $u$ traverses arc $(v_i,v_j)$. Constraints \eqref{energy-constraint} ensure that the total energy consumed by both operations and flight does not exceed the capacity $E_{max}$ for each UAV. Constraints \eqref{total-load} require that the total seed weight $Q$ carried by each UAV must be fully dispensed before returning to the base station. Constraints \eqref{carry-load} govern the payload dynamics: the seed weight carried by each UAV decreases by exactly the amount required at each restored region, while simultaneously eliminating illegal subtours. Constraints \eqref{load} guarantee that the seed demand at each restoration region $v_j$ does not exceed the remaining payload capacity of the servicing UAV. Constraints \eqref{enter-point} ensures that each UAV visits each restoration region at most once and departs after completing the seeding operation. Constraints \eqref{start-point}  requires that each UAV route begins and terminates at the base station. Constraint \eqref{binary-variable} enforces binary integrality on the decision variables. Constraint \eqref{area-constraint} limits the restoration work at each region to not exceed its maximum capacity. Constraint \eqref{flight-weight} imposes nonnegativity restrictions on all relevant variables.

The optimization problem~\eqref{GRP}  is a complex, NP-hard VRP variant with variable demands and nonlinear costs.  Its direct solution is intractable due to three intrinsic challenges: (1) \emph{coupled decision variables}, as restoration region size, seed demand, and UAV trajectories are strictly interdependent; (2) \emph{dynamic problem structure}, where payload-dependent energy consumption fluctuates as service sequences evolve; and (3) \emph{ high computational complexity}, making exact optimization methods infeasible for large-scale instances.

\section{Methodology} \label{Methodology}
This section outlines the problem decomposition and challenges of the RAMP for multi-UAV collaboration grassland restoration. It then introduces the BFPRL for single-UAV trajectory planning and local restoration decisions, and finally details the KC-BFPRL for solving the large-scale RAMP.
\begin{figure*}[htb]
\begin{center}
$\begin{array}{l}
\includegraphics[width=5.0in]{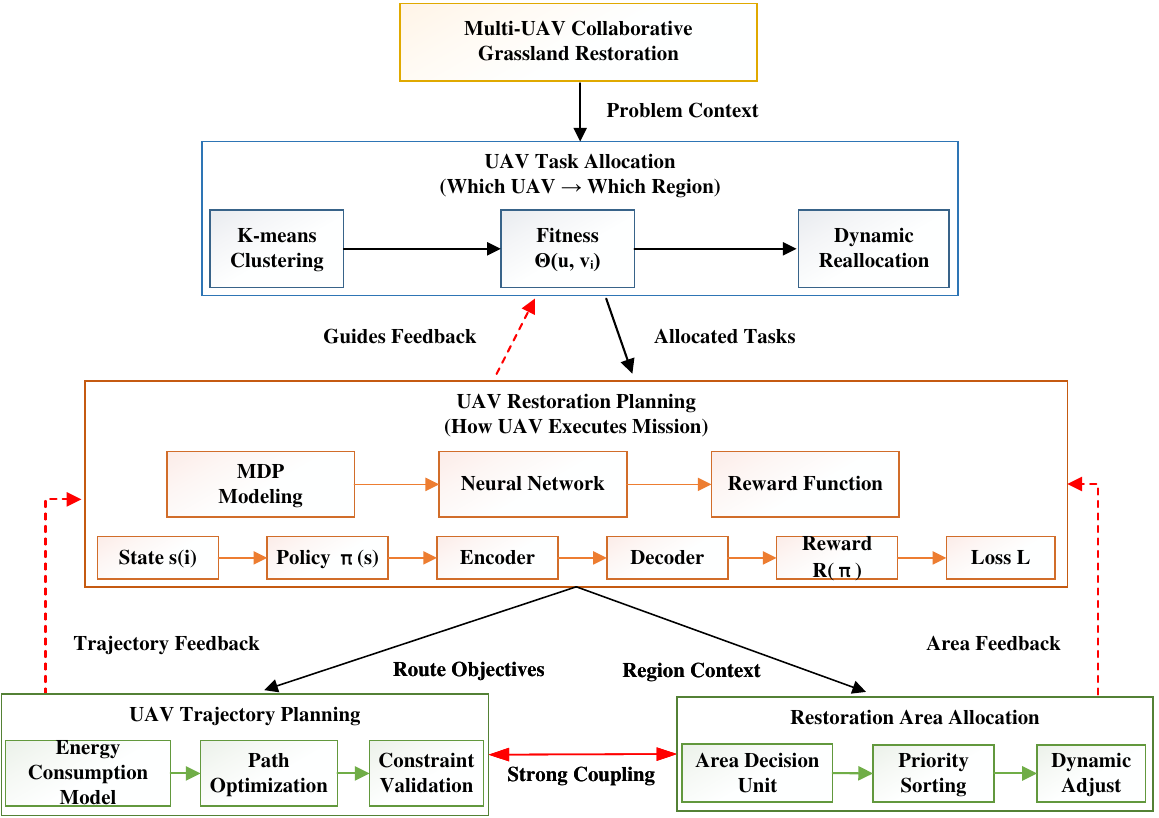}\\
 \end{array}$
\end{center}
\vspace{-0.15in}
\caption{Knowledge-guided multi-UAV collaborative framework.} \label{Knowledge-guided-framework}
\end{figure*}
\subsection{Problem Decomposition}
As illustrated in \ref{Knowledge-guided-framework}, the large-scale problem is decomposed into two interrelated subproblems: multi-UAV task allocation and single-UAV restoration planning. The former assigns specific regions to individual UAVs to balance fleet workloads and maximize resource efficiency. The latter optimizes the trajectory and operational coverage of each UAV, maximizing the total restored area while minimizing energy consumption.  These two subproblems are solved through a hierarchical and collaborative mechanism. The task allocation layer provides the structural framework that guides subsequent restoration planning. Reciprocally, the outcomes of the planning layer dynamically inform and update the task allocation, establishing a real-time feedback loop that ensures the mission is executed cooperatively.

Furthermore, the single-UAV restoration planning subproblem is further decomposed into a tightly coupled bilevel structure: upper-level trajectory planning and lower-level restoration area allocation.  This bilevel coupling is detailed in Section~\ref{methmodel} and is consistent with structures discussed in existing literature~\cite{JIAO2024108084}.
\subsection{Single-UAV Trajectory Planning and Restoration Area Allocation}
\subsubsection{Modeling RAMP via BFPRL}
Modeled as an intelligent agent starting from depot $v_0$, each UAV employs a stochastic policy $\pi_\theta$ to generate a trajectory  
$\tau = \{(v^i_t, a^i_t)\}^T_{t=0}$, where $v^i_t$ and $a^i _t$ denote the visited node and restored area at step $t$, respectively. To maximize the total restored areas under operational constraints, we formulate this process as a four-component MDP and train $\pi_\theta$ using the REINFORCE algorithm with a Greedy Rollout Baseline~\cite{bello2017neural}  to maximize the expected objective in Eq.~\eqref{objectivefunction}.   

\paragraph{State} At each step $t$, the MDP state is defined as $\mathbf{s}_t = (\mathbf{v}_t, q_t^{\mathrm{rem}}, E_t^{\mathrm{rem}}, \mathbf{m}_t, \boldsymbol{\xi}_t)$, where $q_t^{\mathrm{rem}}$ and $E_t^{\mathrm{rem}}$ denote the remaining payload and energy, $\mathbf{m}_t \in \{0,1\}^{N}$ is the visited-node mask vector, and $\boldsymbol{\xi}_t$ tracks the remaining restorable 
unit circles per region. Each node $\mathbf{v}^i_t = (x_i,y_i,l_i)$ contains  its static geographical coordinates $(x_i,y_i)$ and degradation level $l_i$. To avoid ambiguity, action variables are not part of the state, and they are denoted separately in the Action paragraph.
In the BFPRL model, the UAV's state at decision step $i$ is denoted as $\mathbf{s}_{< i}$, where $i \in [1, n + 1]\cap \mathbb{N}^+$. The initial state is $\mathbf{s}_{< 1}=(v_0, Q, E_{\max}, \mathbf{m}_0, \boldsymbol{\xi}_0)$, where only the depot is marked as visited in $\mathbf{m}_0$. Termination occurs when the UAV returns to the depot or no feasible actions remain.

\paragraph{Action} In BFPRL, the action at step $i$ expands the partial solution $\mathbf{s}_{< i}$ by selecting $a_i=(j_i,\delta_i)$, where $j_i$ is the next selected node and $\delta_i$ is the seeding amount (restored circles) executed at $j_i$. To ensure operational viability, a feasibility mask $\mathcal{M}_i(j,\delta) \in \{0,1\}$ restricts sampling to unvisited nodes that satisfy all payload and energy constraints, including a safe return to the depot. This stepwise formulation enables the joint optimization of trajectory planning and area allocation.

\paragraph{Transition} State transitions are deterministic, i.e., $\mathbf{P}\big(s_{< i+1} \mid s_{< i}, s_i\big) = 1$. This means that selecting action $a_i=(j_i,\delta_i)$ in state $\mathbf{s}_{< i}$ transitions the system to the next state $\mathbf{s}_{< i+1}$ with probability 1. Crucially, the transition also includes strict resource updates: remaining energy becomes $E^{\mathrm{rem}}(i+1) = E^{\mathrm{rem}}(i)- E^c(i)$, where the total energy consumption at step $i$ is equal to $E^f(i) + E^s(i) + E^{\mathrm{ap}}(i)$. Meanwhile, the payload becomes $q_{i+1}^{\mathrm{rem}}=q_i^{\mathrm{rem}}-\delta_i q_{j_i}$. The visitation mask and remaining restorable areas are updated via $\mathbf{m}_{i+1}[j_i]=1$ and $\xi_{i+1}(j_i)=\max\{0,\xi_i(j_i)-\delta_i\}$. Infeasible actions are masked to $-\infty$ prior to sampling. If an infeasible action is still selected due to numerical instability, the episode is immediately terminated with a penalty and a forced return to the depot.

\paragraph{Reward} 
Guided by this principle and the specific characteristics of the RAMP, we design a multi-component reward function $R(\pi \mid V)$. This function balances the dual objectives of restoration area maximization and operational energy constraints, defined as follows:
\begin{equation} \label{rewardfunction}
R(\pi \mid V) = \alpha_r \cdot \sum_{i=1}^{n} \big( 1 + (l_i - 0.3) \big) \sigma_i - \alpha_p \cdot p_e,
\end{equation}
where $\alpha_r$ and $\alpha_p$ are balancing coefficients. The penalty $p_e$ is triggered if the energy constraint is violated:
\begin{equation} \label{eq-penalty}
p_e =
\begin{cases}
\sum_{i=0}^{n-1} d_{i,i+1} + d_{n,0}, & E_{\mathrm{rest}} < 0\\[2mm]
0, & E_{\mathrm{rest}} \ge 0,
\end{cases}
\end{equation}
where $d_{i,i+1}$ denotes the flight distance between consecutive restored regions $v_i$ and $v_{i+1}$. If the energy constraint is violated, the penalty equals the total path length. This design encourages the model to prioritize shorter, feasible paths during early training stages while optimizing restoration allocation.

\paragraph{Policy}The restoration process is modeled as a sequence of decisions, where the action at step $i$ is $a_i=(j_i,\delta_i)$ and the policy is factorized via the chain rule:
\begin{equation} \label{policyfunction}
p(\pi \mid s) = \prod_{i=1}^{n} p\big(j_i \mid s_{<i}\big)\, p\big(\delta_i \mid j_i, s_{<i}\big),
\end{equation}
where $p\big(j_i\mid s_{<i}\big)$ denotes the probability of selecting the next node, and $p\big(\delta_i \mid j_i, s_{<i}\big)$ denotes the conditional probability of selecting the restoration level after node $j_i$ is chosen. This factorization explicitly captures the sequential dependency of the joint action on the historical trajectory of visited states and actions.
\subsubsection{BFPRL Architecture Design}
\begin{figure}[htb]
\begin{center}
$\begin{array}{l}
\includegraphics[width=2.6in]{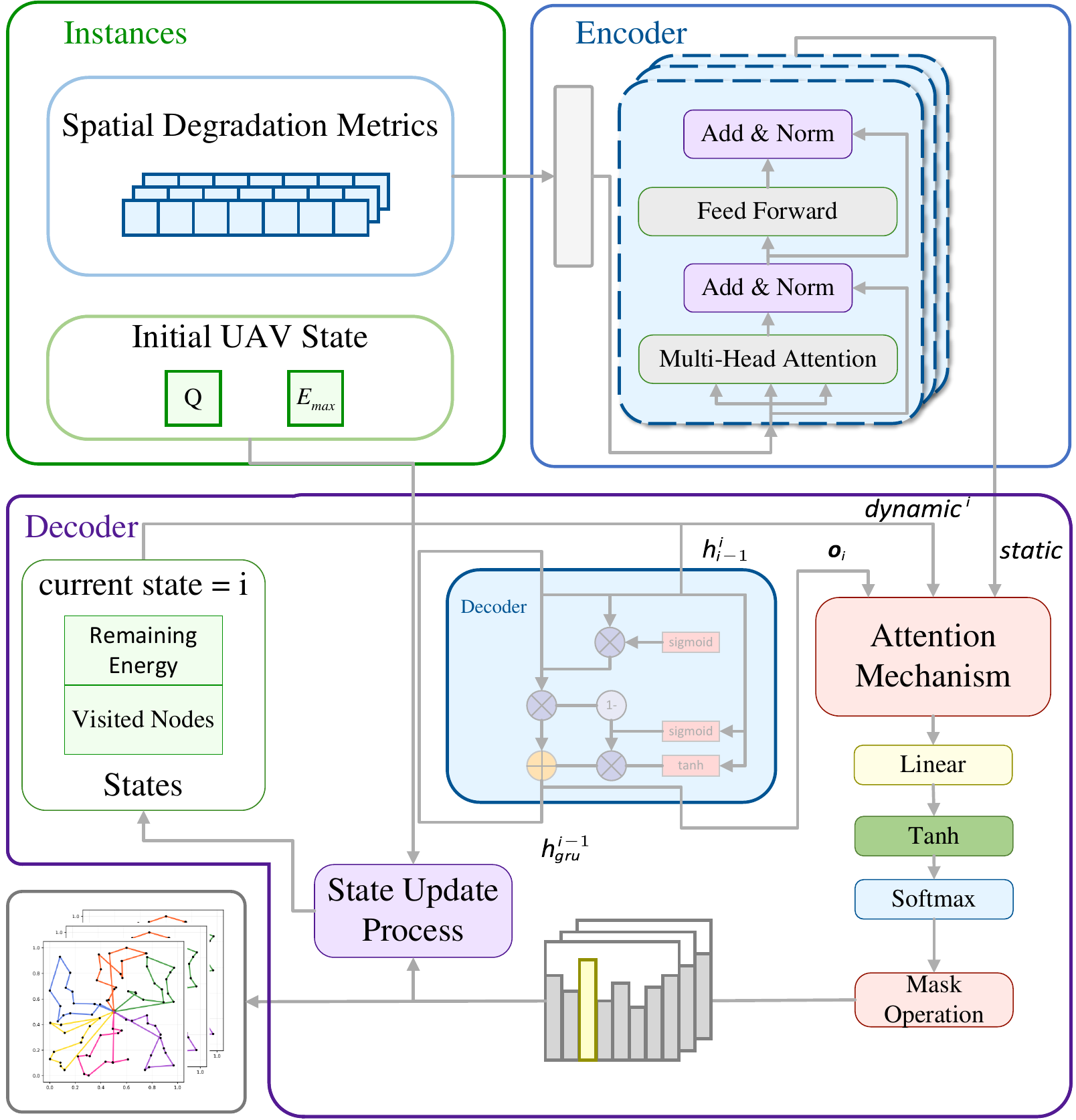}\\
 \end{array}$
\end{center}
\vspace{-0.15in}
\caption{Architecture of BFPRL.} \label{BFPRL-Architecture}
\end{figure}

To learn the stochastic policy $\pi$ in Eq.~\eqref{policyfunction}, we parameterize it as a neural network $\pi_{\theta}$, where $\theta \in \Theta$ denotes the complete set of trainable parameters across the encoder and decoder modules, as shown in Fig. \ref{BFPRL-Architecture}.

\paragraph{State Feature Extraction}
The model $M_{\theta}$ processes both static and dynamic state features.
Static features encompass intrinsic environmental attributes, such as the spatial coordinates and degradation levels of target regions. In contrast, dynamic features capture the time-varying operational status of the UAV, such as remaining energy and payload weight. Optimizing $\Theta$ enables the model to perform high-dimensional feature embedding and sequential decision modeling based on these extracted state representations.

\paragraph{Improved Encoder-Decoder Architecture}
Building on~\cite{bresson2021transformer}, we propose a tailored encoder-decoder architecture for the RAMP.  As illustrated in Fig. \ref{BFPRL-Architecture}, the model adopts a Transformer-based encoder~\cite{vaswani2017attention} to extracts global contextual representations from input sequences. Subsequently, an Pointer Network~\cite{vinyals2015pointer} is integrated as an autoregressive decoder to sequentially generate environment-dependent restoration decisions. This enables the joint optimization of dynamic UAV trajectories and adaptive area allocation.

\paragraph{Encoder}
The encoder integrates heterogeneous inputs, including static environmental features (coordinates, degradation) and dynamic UAV states (energy, payload), to generate a global contextual representation. It consists of 
$n_{\text{layers}}$ Transformer blocks utilizing Batch Normalization (BN) instead of standard LayerNorm to stabilize training.

First, input features $\mathbf{H}_{in} \in \mathbb{R}^{n \times d_{in}}$ are linearly embedded into dimension $d_e$:
\begin{equation}
\mathbf{H}^{l=0} = \mathbf{H}_{in} \mathbf{W}_{in} \in \mathbb{R}^{n \times d_e},\label{eq-2-5}
\end{equation}
where $\mathbf{W}_{in}$ is the input embedding weight matrix. Each subsequent layer $l \in \{0, \ldots, n_{\text{layers}}-1\}$ applies multi-head attention (MHA) and a feed-forward network (FFN) with residual connections and BN: 
\begin{align}
\mathbf{H}^{l+1}_{rc} &= \text{BN}\left(\text{MHA}^{l+1}(\mathbf{H}^l) + \mathbf{H}^l\right), \quad \\
\mathbf{H}^{l+1} &= \text{BN}\left(\text{ReLU}\left(\mathbf{H}^{l+1}_{rc}\mathbf{W}_1^{l+1}\right)\mathbf{W}_2^{l+1} + \mathbf{H}^{l+1}_{rc}\right), \quad
\end{align}
where $\text{MHA}^{l+1}(\cdot)$ denotes the attention mechanism at layer $l+1$, $\mathbf{H}^{l+1}_{rc}$ is the intermediate residual connection, $\mathbf{W}_1^{l+1}$ and $\mathbf{W}_2^{l+1}$ are the weight matrices of the FFN, and $\text{BN}(\cdot)$ denotes batch normalization. The final output  $\mathbf{H}^{(n_{\text{layers}})}$ provides the comprehensive global representation required by the decoder.

\paragraph{Decoder}
The autoregressive decoder constructs the UAV restoration plan via stepwise decoding. At step $i$, a Gated Recurrent Unit (GRU) maintains the sequential hidden state:
\begin{equation}
\mathbf{o}_i, \mathbf{H}_i^{\text{gru}} =
\begin{cases}
\text{GRU}(\mathbf{H}_i^0, \mathbf{0}), & i = 0 \\
\text{GRU}(\mathbf{H}_i^{i-1}, \mathbf{H}_{i-1}^{\text{gru}}), & i > 0,
\end{cases}
\end{equation}
where $\mathbf{H}_i^{\text{gru}}$ is the GRU hidden state, $\mathbf{o}_i$ is the GRU output, $\mathbf{H}_i^0$ is the initial input, and $\mathbf{H}_i^{i-1}$ is the current input feature at step $i$. This recurrent mechanism ensures that the decision at the current step is informed by the complete historical trajectory constructed so far.

Conditioned on the encoder output and the current dynamic state, an attention mechanism computes the decoder context representation at $i$ as follows:
\begin{equation}
\mathbf{h}^{\text{dec}}_i = \text{Attention}\left(\mathbf{H}^{(n_{\text{layers}})}, \text{dynamic}_i, \mathbf{o}_i\right).
\end{equation}

Following the computation of the contextual representation, the pointer decoder evaluates the selection probability for each candidate node. The raw selection score $\mu_{ij}$ for candidate node $j$ is computed as:
\begin{equation}
\mu_{ij} =
\begin{cases}
\zeta \cdot \tanh \left((\mathbf{W}_q\mathbf{h}^{\text{dec}}_i)^\top (\mathbf{W}_k\mathbf{h}^{\text{enc}}_j)\right), & j \notin \pi_i \\
-\infty, & \text{otherwise},
\end{cases}
\end{equation}
where $\mathbf{W}_q$ and $\mathbf{W}_k$ are query and key weight matrices, and $\zeta$ is a scaling factor (typically set to 10). If $\pi_i$ is the set of already visited nodes, its score is set to $-\infty$, which ensures that the node cannot be selected again. The node-selection probability is then obtained by
\begin{equation}
p_\theta(j_i=j \mid s_{<i}) = \text{Softmax}(\mu_i)_j.
\end{equation}

After the node $j_i$ is selected, an area-allocation head predicts the restoration level from a feasible set $D_i(j_i)$. This set is dynamically constrained by the remaining payload, energy, and residual restorable areas:
\begin{equation}
\nu_{i\delta}^{(j_i)} = \mathbf{w}_{\delta}^{\top}\mathrm{ReLU}\left(\mathbf{W}_{\delta}\left[\mathbf{h}^{\text{dec}}_i;\mathbf{h}^{\text{enc}}_{j_i};\text{dynamic}_i\right]\right), \quad \delta \in D_i(j_i),
\end{equation}
\begin{equation}
p_\theta(\delta_i=\delta \mid j_i, s_{<i}) = \text{Softmax}\left(\nu_i^{(j_i)}\right)_{\delta}.
\end{equation}
Accordingly, the joint action distribution factorizes as
\begin{equation}
p_\theta(a_i \mid s_{<i}) = p_\theta(j_i \mid s_{<i})\, p_\theta(\delta_i \mid j_i, s_{<i}).
\end{equation}

During forward propagation, the model $M_\theta(x)$ takes the input problem instance $x$ and
outputs both the trajectory cost $C$ and the aggregated log-probability:
\begin{equation}
(C, \log p) \leftarrow M_\theta(x),
\end{equation}
where the aggregated log-probability across all $L$ steps is given by:
\begin{equation}
\log p = \sum_{i=1}^{L} \left[\log p_\theta(j_i \mid s_{<i}) + \log p_\theta(\delta_i \mid j_i, s_{<i})\right].
\end{equation}
This trajectory-level output facilitates the subsequent gradient-based optimization of the policy parameters $\theta$.

\subsubsection{Model Training and Optimization}
To train the single-UAV restoration model, we employ an actor-critic algorithm~\cite{sutton1999policy} combined with the Adam optimizer \cite{kingma2015adam} to update the policy network $\pi_\theta$. As an extension of the REINFORCE algorithm~\cite{williams1992simple}, this dual-network approach effectively stabilizes training and accelerates convergence.

\paragraph{Actor Network Optimization}
The actor network approximates the policy probability function $p(\pi|s)$ for the restoration task. Let $\theta$ denote its trainable parameters, the optimization objective is to maximize the expected return under the policy induced by $\theta$:
\begin{equation}
J(\theta \mid s) = \mathbb{E}_{\pi \sim p_{\theta}(\cdot \mid s)}[R(\pi \mid s)].
\end{equation}

Following~\cite{williams1992simple}, the policy gradient is derived using the advantage function $A(\pi \mid s)$:
\begin{equation}
\begin{aligned}
\nabla_{\theta} J(\theta \mid s)
&= \mathbb{E}_{\pi \sim p_{\theta}(\cdot \mid s)}
\left[ \left( R(\pi \mid s) - b(s) \right)
\nabla_{\theta} \ln p_{\theta}(\pi \mid s) \right] \\
&= \mathbb{E}_{\pi \sim p_{\theta}(\cdot \mid s)}
\left[ A(\pi \mid s)
\nabla_{\theta} \ln p_{\theta}(\pi \mid s) \right],
\end{aligned}
\end{equation}
where $R(\pi \mid s)$ denotes the reward obtained during the grassland restoration process, and $p_{\theta}(\pi \mid s)$ is the probability that the actor network selects a restoration region and its corresponding restoration area size at each autoregressive decision step. The term $b(s)$ is the baseline value estimate provided by the critic network.  The advantage function $A(\pi \mid s) = R(\pi \mid s) - b(s)$ evaluates the relative performance of policy $\pi$ against the expected value of state $s$. A positive advantage reinforces the selected actions via the term $\nabla_{\theta} \ln p_{\theta}(\pi \mid s)$, whereas a negative advantage reduces suboptimal actions.

For a training batch of size $B$ with $L$
decision steps, the gradient is approximated as:
\begin{equation}
\nabla_{\theta} J(\theta \mid s) \approx \frac{1}{B} \sum_{i=1}^{B} \sum_{j=1}^{L} \left[ A(\pi_{i,j} \mid s_{i,j}) \nabla_{\theta} \ln p_{\theta}(\pi_{i,j} \mid s_{i,j}) \right].
\end{equation}

\paragraph{Critic Network Optimization}
Let $\theta_c$ denote the parameters of the critic network, the objective of the critic is to minimize the discrepancy between the predicted baseline value $b_{\theta_c}(V_i)$ and the experienced return $R(\pi \mid V_i)$. Accordingly, the loss function can be approximated by the mean squared error (MSE):
\begin{equation}
\mathcal{L}(\theta_c \mid s) \approx \frac{1}{B} \sum_{i=1}^{B} \left\| b_{\theta_c}(V_i) - R(\pi \mid V_i) \right\|_2^2.
\end{equation}

\paragraph{Unified Training Procedure}
\begin{algorithm}[!htbp] \small
\caption{BFPRL Training for Single-UAV Trajectory Planning and Restoration Area Allocation}
\label{BFPRL-Framework}
\SetKwInOut{KwIn}{Input}
\SetKwInOut{KwOut}{Output}

\KwIn{Initial model parameters $\theta$, training epochs $N_{\text{epoch}}$, batch size $B$, learning rate $\alpha$, gradient clipping threshold $g_{\max}$, validation set size $|D_{\text{val}}|$.}
\KwOut{Optimal policy parameters $\theta^*$, trained actor network $\mathcal{M}_{\theta^*}$, critic baseline network $\mathcal{B}^*$.}

\tcp{Initialization}
$\mathcal{M}_\theta \leftarrow \text{InitializeActor}(\theta)$; \quad $\mathcal{B}_{\text{model}} \leftarrow \text{InitializeCritic}()$\;
$\mathcal{O} \leftarrow \text{Adam}(\{\mathcal{M}_\theta, \mathcal{B}_{\text{model}}\}, \alpha)$ \tcp*[r]{Init Optimizer}
$D_{\text{val}} \leftarrow \text{GenerateInstances}(|D_{\text{val}}|)$ \tcp*[r]{Validation Set}
$R_{\text{best}} \leftarrow -\infty$; \quad $\theta^* \leftarrow \theta$\;

\tcp{Main Training Loop}
\For{epoch $\leftarrow 1$ \KwTo $N_{\text{epoch}}$}{
    $D_{\text{train}} \leftarrow \text{GenerateInstances}()$ \tcp*[r]{Training Set}

    \ForEach{Batch $\in \text{Split}(D_{\text{train}}, B)$}{
        $(C, \log p) \leftarrow \mathcal{M}_\theta(\text{Batch})$ \tcp*[r]{Forward pass}
        $b_v \leftarrow \mathcal{B}_{\text{model}}(\text{Batch})$ \tcp*[r]{Baseline estimate}

        \tcp{Compute Losses (Eq. 26)}
        $\mathcal{L}_R \leftarrow \frac{1}{B} \sum_{i=1}^B (C_i - b_{v,i}) \cdot \log p_i$\;
        $\mathcal{L}_b \leftarrow \text{BaselineLoss}(b_v, C)$\;
        $\mathcal{L} \leftarrow \mathcal{L}_R + \mathcal{L}_b$ \tcp*[r]{Total Loss}

        \tcp{Backpropagation}
        $\nabla_\theta \leftarrow \nabla_\theta \mathcal{L}$\;
        $\|\nabla_\theta\|_2 \leftarrow \min(\|\nabla_\theta\|_2, g_{\max})$ \tcp*[r]{Gradient Clipping}
        $\theta \leftarrow \mathcal{O}(\theta, \nabla_\theta)$ \tcp*[r]{Update Parameters}
    }

    \tcp{Validation Phase}
    $\mathcal{M}_\theta.\text{SetMode}(\text{GreedyDecoding})$\;
    $C_{\text{val}} \leftarrow \text{EvaluatePolicy}(\mathcal{M}_\theta, D_{\text{val}})$\;
    $\bar{r} \leftarrow -\frac{1}{|D_{\text{val}}|} \sum_{i=1}^{|D_{\text{val}}|} C_{\text{val}, i}$ \tcp*[r]{Avg Reward}

    \If{$\bar{r} > R_{\text{best}}$}{
        $\theta^* \leftarrow \theta$; \quad $R_{\text{best}} \leftarrow \bar{r}$ \tcp*[r]{Save Best Model}
    }
}
\Return $\theta^*, \mathcal{M}_{\theta^*}, \mathcal{B}^*$\;
\end{algorithm}

As detailed in Algorithm~\ref{BFPRL-Framework}, the actor and critic networks are jointly optimized using Adam with a scheduled learning rate $\alpha$. The total loss $\mathcal{L} = \mathcal{L}_R + \mathcal{L}_b$ combines the critic's baseline loss $\mathcal{L}_b$ and the actor's policy gradient loss:
\begin{equation}\label{Policy-gradient-loss}
\mathcal{L}_R = \frac{1}{B} \sum_{i=1}^{B} (C_i - b_{v,i}) \cdot \log p_i,
\end{equation}
where $C_i$ is the actual cost (negative reward) and $b_{v,i}$ is the critic-estimated baseline. To prevent gradient explosion, gradients are clipped via $\|\nabla_{\theta}\|_2 \leftarrow \min(\|\nabla_{\theta}\|_2, g_{\max})$ prior to parameter updates. After each epoch, the model is evaluated on a validation set $D_{\text{val}}$ via greedy decoding. We retain the optimal parameters $\theta^*$ that maximize the average validation reward:
\begin{equation}
\theta^* = \arg\max_{\theta} \left( -\frac{1}{|D_{\text{val}}|} \sum_{i=1}^{|D_{\text{val}}|} C_{\text{val},i} \right)
\end{equation}

This architecture ensures stable gradient estimation and rapid convergence to an optimal single-UAV restoration policy.

\subsection{Knowledge-Guided Multi-UAV Collaboration Framework}
To address the challenges of large-scale degraded grassland restoration, we propose KC-BFPRL, a knowledge-guided multi-UAV collaboration framework. Built upon a bilevel formerpointer reinforcement learning approach, KC-BFPRL adopts the hierarchical collaborative architecture illustrated in Fig. \ref{Knowledge-guided-framework}.
Following an ``external collaboration, internal intelligence" paradigm, a ground control center (GCC) manages global cooperative scheduling and dynamic map updates.  Concurrently, individual UAVs utilize the Transformer-Pointer based BFPRL model for local sequential decision-making to optimize flight trajectories and restoration allocations. This architecture strategically couples global scheduling data with local perception, effectively balancing computational efficiency and global restoration performance.

\subsubsection{Collaboration Scheduling Mechanism}
To facilitate collaboration, the global target set $V_a = V \setminus \{v_0\}$ is partitioned into $m$ mutually exclusive subsets, one for each UAV $u \in U = \{1,\ldots, m\}$, satisfying $m \ll n$.
Each subset $V_u = \{v^1_u, v^2_u, \dots, v^h_u\}$ assigns $h$ regions to UAV $u$, where each region is defined by a tuple $v_u^i = (\mathrm{id}_u^i, p_u^i, l_u^i, a_u^i)$ representing its identifier, spatial coordinates, degradation level, and restorable area size, respectively.

The operational state of UAV $u$ is defined as $S_u = \left( \mathbf{p}_u, E_u^{\text{rem}}, V^{\text{vis}}_u, A^{\text{rep}}_u \right)$, where $\mathbf{p}_u$ is the current position, $E^{\text{rem}}_u$ is the remaining energy, $V^{\text{vis}}_u$ tracks the sequence of visited regions, and $A^{\text{rep}}_u$ is the total accumulated restored areas.

To synchronize interactions between the UAVs and the GCC, a semaphore variable $SP_u$ is introduced to estimate the time cost for the next task:
\begin{equation}
SP_u = \frac{d(\mathbf{p}_u, v_{\text{next}}^i)}{v} + \rho \cdot \frac{a^{\text{rep}}_i}{r},
\end{equation}
where $d(\cdot)$ is the Euclidean distance, $v$ denotes the UAV's flight speed, $r$ represents the restoration rate (areas per unit time), $\rho$ is a weighting factor, and $a_i^{\text{rep}}$ is the target region size at the next node $v^{\text{next}}_i$.

\subsubsection{Multi-UAV Information Sharing Strategy}
To optimize task distribution via dynamic information exchange, a matching cost function
$K(u,v_i)$ evaluates the suitability of assigning region $v_i$ to UAV $u$. Balancing spatial proximity, task urgency, and workload, it is formulated as:
\begin{equation}
K(u, v_i) = \alpha \cdot d(\mathbf{p}_u, \mathbf{p}_i) + \beta \cdot \frac{1}{l_i} + \epsilon \cdot \frac{1}{a_i},
\end{equation}
where $\alpha$, $\beta$, $\epsilon$ are tunable weight coefficients for distance, degradation level, and restoration area size, respectively. A lower $K$ value indicates a higher assignment priority.

To achieve global optimization, this cooperative scheduling process iteratively executes the following five key steps.

\begin{description}
  \item[\textbf{Step 1:}]~\emph{Initial Partitioning}. The global target region $V_a$ is partitioned via spatial clustering (e.g., K-means), allocating an initial subset $V_u^0$ to each UAV $u$. \label{step1}
  \item[\textbf{Step 2:}]~\emph{Baseline Trajectory Optimization}. Based on $V_u^0$, each UAV computes an energy-feasible trajectory to maximize its baseline restoration gain $A_{u1}$:
\begin{equation}
(P_{u1}, A_{u1}) = \arg\max_{P \in \mathcal{P}(M_u)} \sum_{v_i \in P} a_i^{\text{rep}},
\end{equation}
where $\mathcal{P}(M_u)$ denotes the set of energy-feasible paths and $a_i^{rep}$ is the number of restoration areas in region $v_i$.  

\item[\textbf{Step 3:}]~\emph{Global Task Reallocation}. 
  The GCC tentatively reassigns tasks based on the real-time fleet states, allocating each region to the UAV that minimizes the matching cost $K$:
\begin{equation}
V_u^{\text{tmp}} = \left\{v_i \in V_a \middle| u = \arg\min_{u' \in U} K(u', v_i)\right\}.
\end{equation}
  \item[\textbf{Step 4:}]~\emph{Candidate Trajectory Generation}. Based on the tentative partition $V_u^{tmp}$, each UAV replans its trajectory to determine the potential restoration gain, denoted as $A_{u2}$.
  \item[\textbf{Step 5:}]~\emph{Global Decision}. The system evaluates the efficacy of the reallocation by comparing aggregate restoration gains of baseline (Step 2) and candidate (Step 4) plans. The tentative allocation is adopted only if it improves global performance:
    \begin{equation}
    V_u =
    \begin{cases}
    V_u^{\text{tmp}}, & \text{if } \sum_{u \in U} A_{u2} \geq \sum_{u \in U} A_{u1} \\
    V_u, & \text{otherwise}.
    \end{cases}
    \end{equation}
\end{description}

This iterative information-sharing mechanism facilitates dynamic task reallocation, maximizing the total restored area under strict  energy constraints.
\subsubsection{Decision-Making and Execution}
UAVs execute restoration decisions sequentially based on calculated semaphore priorities. The UAV with the minimum semaphore is selected first:
\begin{equation}
u^* = \arg\min_{u \in U} SP_u.
\end{equation}

Upon completing its current task at $v^{\text{curr}}_i$, UAV $u^*$ selects the next target region within its assigned subset by minimizing the local matching cost:
\begin{equation}
v^{\text{next}}_i = \arg\min_{v_i \in V_{u^*}} K(u^*, v_i).
\end{equation}

The state of $u^*$  is then updated as:
\begin{equation}\label{optimal-state}
\begin{split}
S_{u^*} = & \left( v_i^{\text{next}}, \ E_{u^*}^{\text{rem}} - E_{\text{flight}} - E_{\text{repair}},\right. \\
& \left. V_{u^*}^{\text{vis}} \cup \{v_i^{\text{curr}}\}, \ A_{u^*}^{\text{rep}} + a_i^{\text{rep}} \right),
\end{split}
\end{equation}

The process (summarized in Algorithm~\ref{alg-KC-BFPRL}) repeats until all global tasks are completed or energy constraints force a return to the depot. 
\begin{algorithm}[t] \small
\caption{Knowledge-Guided Multi-UAV Collaboration Scheduling Algorithm for RAMP}
\label{alg-KC-BFPRL}
\SetKwInOut{KwIn}{Input}
\SetKwInOut{KwOut}{Output}

\KwIn{Global parameter set $P$, initial restoration map partitions $\{V_u^0\}_{u=1}^{m}$, UAV states $\{S_u^0\}_{u=1}^{m}$, pre-trained RL model parameters $\mathcal{M}_\theta$.}
\KwOut{UAV trajectories $\{P_u\}_{u=1}^{m}$, cumulative restored areas $\{A_u\}_{u=1}^{m}$, remaining energy $\{E_u\}_{u=1}^{m}$.}

\tcp{Initialization}
Initialize $V_u \leftarrow V_u^0, P_u \leftarrow \emptyset, S_u \leftarrow S_u^0, \forall u \in U$; \quad $\mathcal{M}_{\text{global}} \leftarrow \bigcup V_u$\;

\tcp{Main Scheduling Loop}
\While{$\mathcal{M}_{\text{global}} \neq \emptyset$}{
    \tcp{Phase I: Distributed Planning \& Global Reallocation}
    Compute baseline: $(E_u^{(1)}, A_u^{(1)}) \leftarrow \text{PlanTrajectory}(\mathcal{M}_\theta, V_u) \ \forall u \in U$\;
    Update global map: $\mathcal{M}_{\text{global}} \leftarrow \text{UpdateAllocation}(\{V_u, A_u^{(1)}\})$\;
    Compute candidate: $V_u^{\text{tmp}} \leftarrow \text{GetTasks}(\mathcal{M}_{\text{global}})$; \ $(E_u^{(2)}, A_u^{(2)}) \leftarrow \text{PlanTrajectory}(\mathcal{M}_\theta, V_u^{\text{tmp}}) \ \forall u$\;

    \lIf{$\sum A_u^{(2)} \ge \sum A_u^{(1)}$}{Update maps: $\{V_u\} \leftarrow \{V_u^{\text{tmp}}\}$}

    \tcp{Phase II: Semaphore-based Execution}
    Select active UAV: $u^* \leftarrow \arg\min_{u} SP_u$; \quad Select target: $v^* \leftarrow \arg\min_{v \in V_{u^*}} K(u^*, v)$\;

    \If{$v^* \neq \text{NULL}$}{
        ExecuteRestoration($u^*, v^*$); \quad $V_{u^*} \leftarrow V_{u^*} \setminus \{v^*\}$\;
        Update state $S_{u^*}$ (Energy, Position, Area) per Eq.~\eqref{optimal-state}\;
        Update semaphore $SP_{u^*}$ based on new state\;
    }
}

\tcp{Termination}
\lForEach{UAV $u \in U$}{ReturnToDepot($u$)}\;
\Return $\{\mathcal{P}_u\}_{u=1}^U, \{A_u\}_{u=1}^U, \{E_u\}_{u=1}^U$.
\end{algorithm}
\section{Experimental Evaluation and Analysis} \label{Experimental}

This section presents comprehensive experiments validating the effectiveness and generalization of the proposed KC-BFPRL framework for the RAMP.
\subsection{Experiment Settings}
All experiments were conducted on a workstation equipped with an AMD Ryzen Threadripper 3970X CPU, NVIDIA RTX 3090 Ti GPU, and 192 GB RAM, utilizing Python 3.9 and PyTorch 1.12.1 (CUDA 11.3) on Ubuntu 22.04.
\subsubsection{Instance Generation}
To evaluate the KC-BFPRL framework's generalization across diverse spatial scales and constraints, we generated 10800 test instances spanning 108 parameter configurations. These configurations combine three fleet sizes $M\in \{4,6,8\}$, six problem scales $N\in \{60,80,\ldots,160\}$ target regions, spanning areas from $500 \times 500 \, \text{m}^2$ to $1000 \times 1000 \, \text{m}^2$, and six workload levels, where the number of unit circles $\delta$ per degraded region is uniformly sampled from the set $\{10,15,\ldots,35\}$ in increments of $5$ across all scenarios. Configurations are denoted as UM-RN, e.g., U4-R60.

For each instance, the BS is fixed at $(0,0)$. Target coordinates are uniformly sampled in $[0,1]^2$ and scaled to the specific spatial area, with degradation levels $l_i \sim U(0,1)$. To accommodate larger domains, the initial UAV energy $E_{max}$ is proportionally scaled from $1.00\times 10^7$ J to $7.59 \times 10^7$ J. To ensure statistical reliability, we evaluate 100 independent random instances per configuration.

\subsubsection{Compared Algorithms}
We evaluate KC-BFPRL against three state-of-the-art baselines:
(1)~\textbf{CHAPBILM~\cite{JIAO2024108084}}: A specialized heuristic for UAV grassland restoration combining population-based incremental learning with a maximum-residual-energy local search. (2)~\textbf{MAPDP~\cite{zong2022mapdp}}: A multi-agent routing algorithm utilizing paired context embedding and a collaborative advantage actor-critic (A2C) approach.
(3)~\textbf{CAMP~\cite{hua2025camp}}: An attention-based multi-agent model featuring inter-agent communication and a centralized-training distributed-execution (CTDE) framework.

Since the original \textbf{MAPDP} and \textbf{CAMP} were not originally designed for the RAMP, we adapted their input layers and decoder masking mechanisms to accommodate our problem constraints while preserving their core architectures. Detailed parameter configurations are provided in  Appendix \ref{Hyperparameter}.

\subsubsection{Parameter Settings}
Following~\cite{dorling2017vehicle}, the UAV operational parameters are set as follows: the frame mass is initialized to $M = 1.5$ kg, gravitational acceleration to $g = 9.8$ m/s$^2$, air density $\rho = 1.024$ kg/m$^3$, rotor disc area $\varsigma = 0.2$ m$^2$, and the number of rotors to $h = 6$. Regarding the energy consumption dynamics in Section~\ref{enegymodel}, the seeding energy coefficient is set to $\eta = 5 \times 10^4$, the environmental parameter to $\gamma = 1.5$, and the aerial photography energy consumption to $e^{ap} = 2 \times 10^4$ J. 
\subsection{Performance Evaluation}
Following established evaluation protocols~\cite{JIAO2024108084,mao2024dl}, we evaluate KC-BFPRL framework across six key performance metrics: average objective value (Obj.), average optimality gap (Gap), average energy efficiency (ratio of task energy to total energy), average total trajectory length, average total restoration areas (Areas), and average computation time (Time). All data are rounded to the shown precision (with Areas rounded down), and best results are in bold.
\subsubsection{Scalability and Performance Comparison}
To validate its adaptability and scalability, the KC-BFPRL framework  is evaluated across diverse problem scales. The task complexity is systematically tested using target node counts ranging from $n=60$ to $n=160$ in increments of 20, spanning scenarios from small-to-medium scenarios to medium-to-large scales. For each scale, performance is further assessed under varying fleet configurations of 4, 6, and 8 UAVs.

\begin{figure}[htbp]
\begin{center}
$\begin{array}{l}
\includegraphics[width=3.0in]{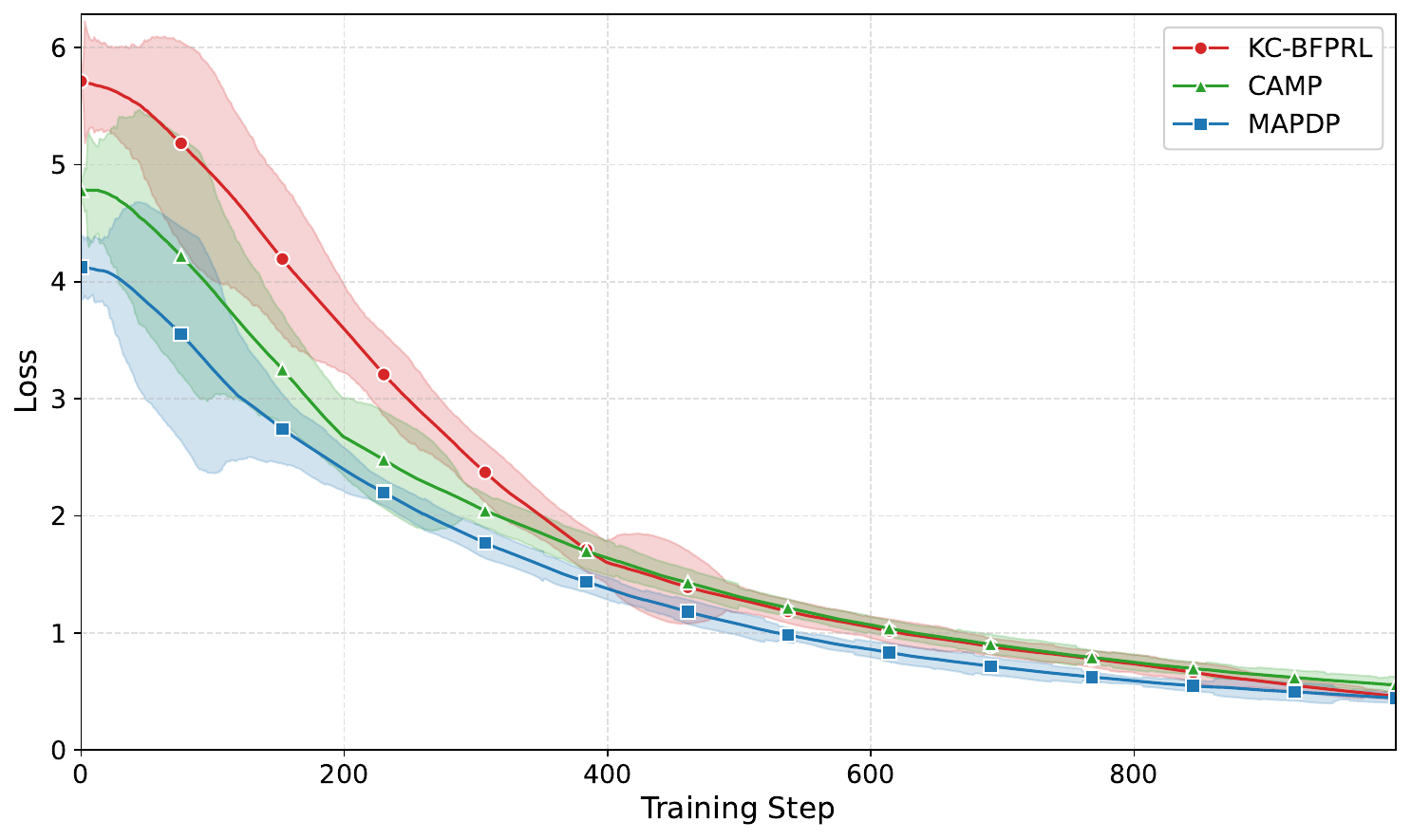}\\
 \end{array}$
\end{center}
\vspace{-0.15in}
\caption{Comparison of training convergence curves for the three learning-based algorithms in scenario U4-R60.} \label{Training-Loss-Comparison}
\end{figure}

First, we evaluate training stability of the model in Fig. \ref{Total-Area-Training-Episodes} of Appendix \ref{extended-experimental}, which reveals a trade-off between efficiency and performance.  A distinct trade-off between training efficiency and restoration performance is evident. While MAPDP demonstrates the fastest convergence and CAMP converges moderately with limited improvement in restoration area, KC-BFPRL requires the longest training period. However, this extended training enables KC-BFPRL to achieve a significantly superior restoration outcome. As confirmed by the final convergence states in Fig. \ref{Training-Loss-Comparison}, the KC-BFPRL models exhibit robust stability, indicating the successful acquisition of effective, high-performance restoration policies. Subsequently, to provide intuitive insight into the model's decision-making, Fig. \ref{Optimal-KC-BFPRL-trajectories-small-to-medium-scale} of Appendix \ref{extended-experimental} visualizes the optimal trajectories for random instances with seed 1234. These results confirm that KC-BFPRL generates logical, efficient paths across various scales, effectively coordinating the multi-UAV fleet to maximize restoration coverage.
\subsubsection{Comparison With State-of-the-Art Methods}
\begin{figure}[htb]
    \centering
    \subfigure[U4-R60]{\includegraphics[width=0.8in]{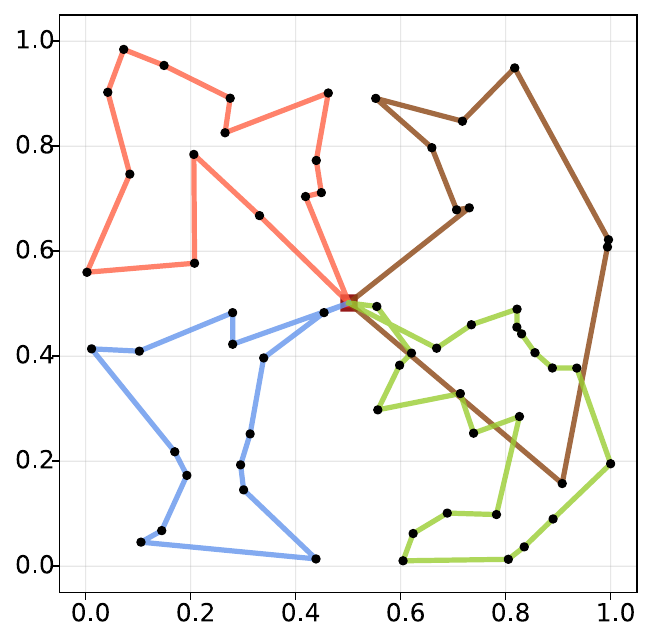}}
    \subfigure[U6-R80]{\includegraphics[width=0.8in]{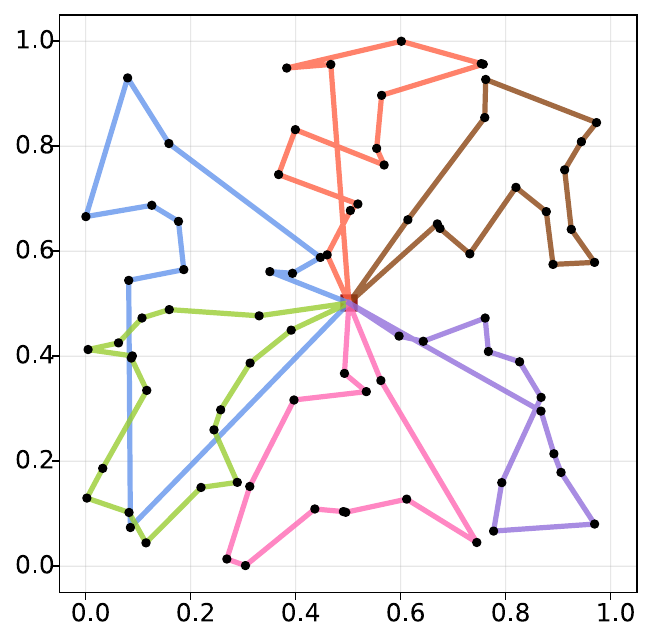}}
    \subfigure[U8-R100]{\includegraphics[width=0.8in]{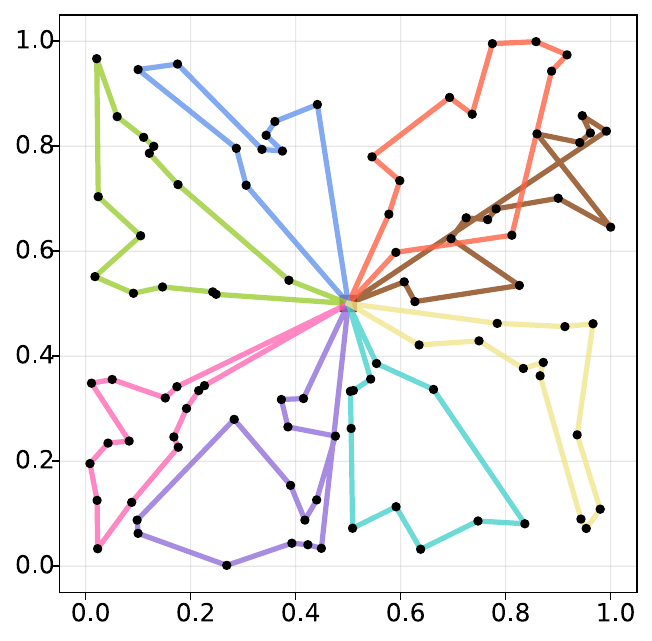}}
    \subfigure[U4-R60]{\includegraphics[width=0.8in]{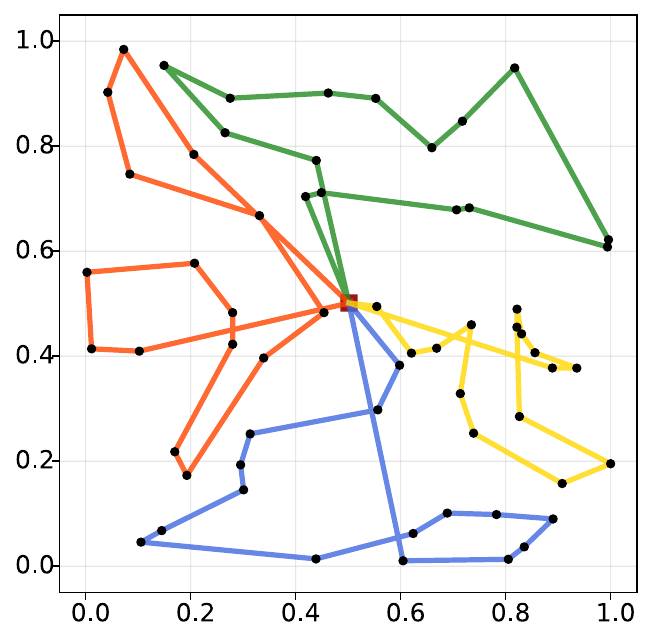}}
    \subfigure[U6-R80]{\includegraphics[width=0.8in]{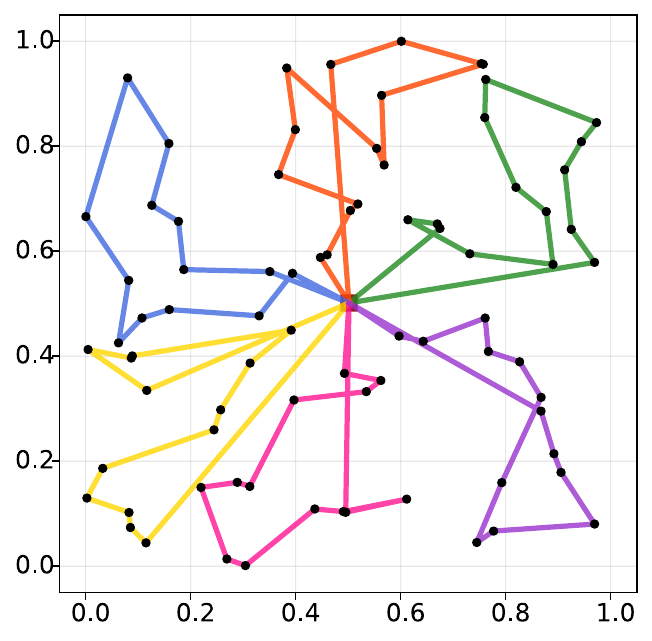}}
    \subfigure[U8-R100]{\includegraphics[width=0.8in]{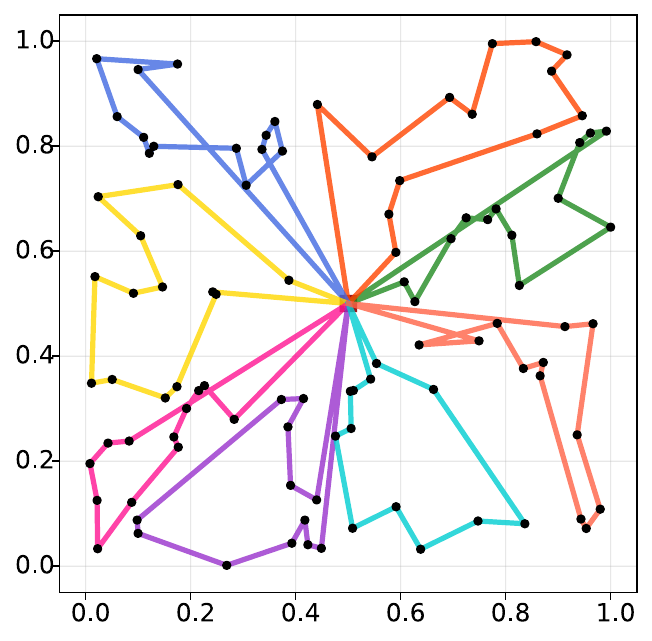}}
    \subfigure[U4-R60]{\includegraphics[width=0.8in]{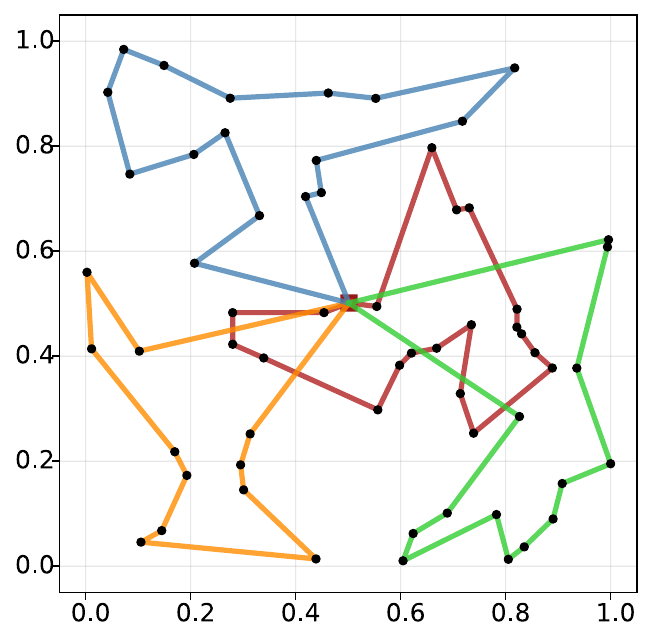}}
    \subfigure[U6-R80]{\includegraphics[width=0.8in]{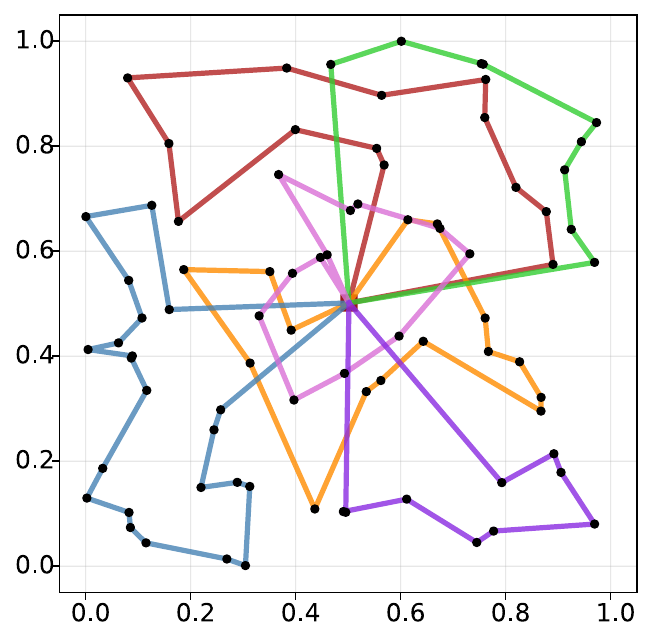}}
    \subfigure[U8-R100]{\includegraphics[width=0.8in]{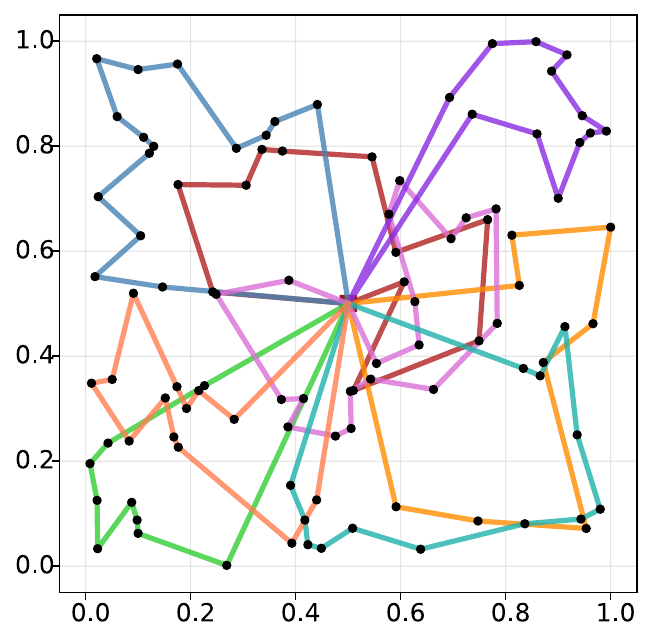}}
    \subfigure[U4-R60]{\includegraphics[width=0.8in]{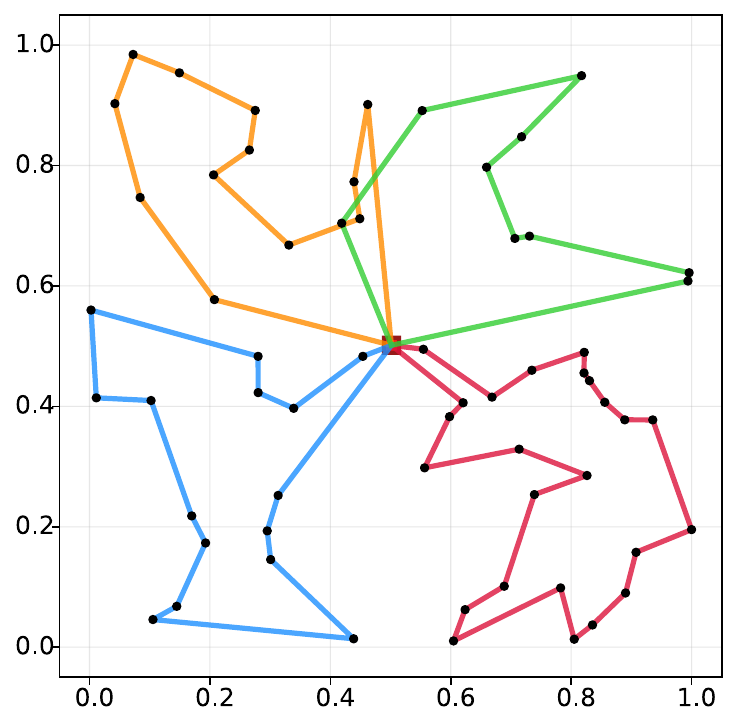}}
    \subfigure[U6-R80]{\includegraphics[width=0.8in]{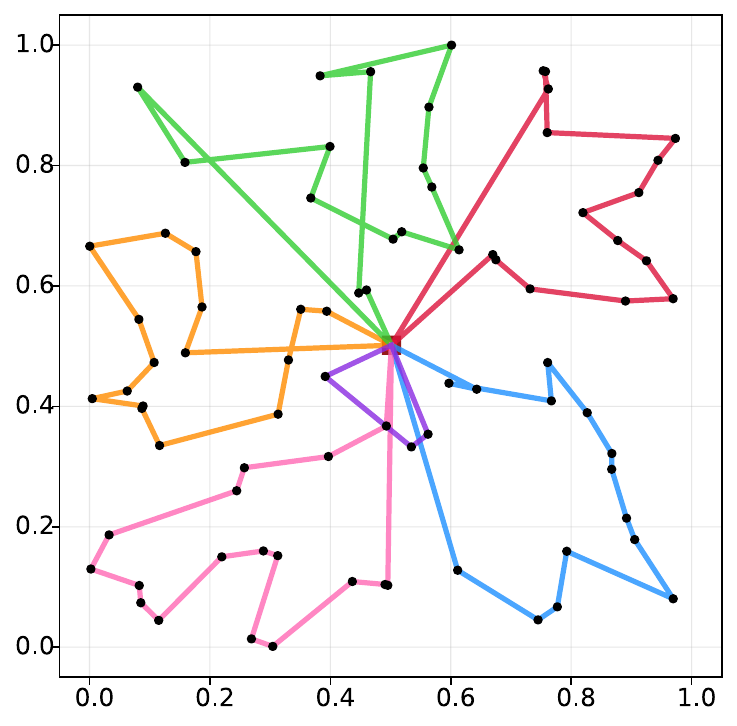}}
    \subfigure[U8-R100]{\includegraphics[width=0.8in]{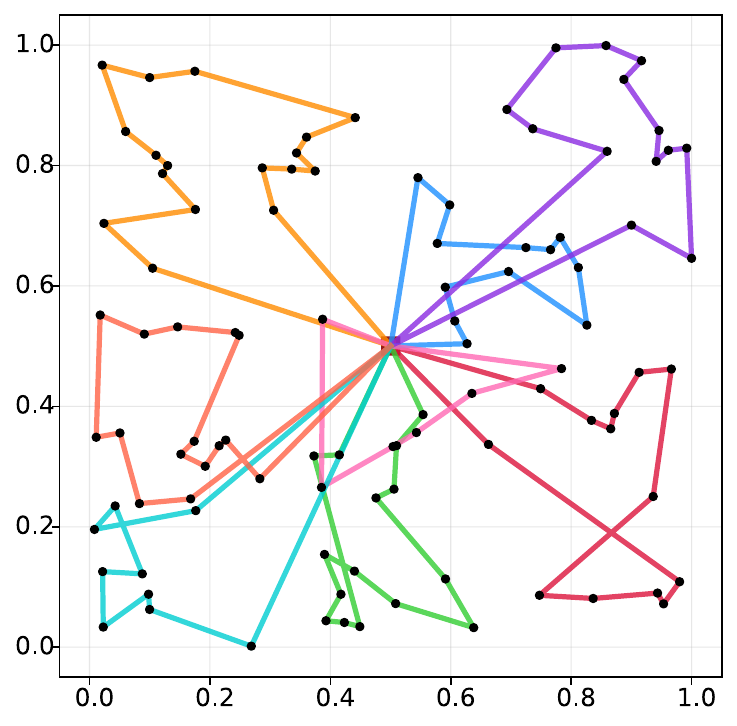}}
\caption{Optimal trajectories generated by the four algorithms for 4, 6, and 8 UAVs across three small-to-medium-scale scenarios: (a)-(c) CAMP, (d)-(f) CHAPBILM, (g)-(i) MAPDP, and (j)-(l) KC-BFPRL.} \label{Optimal-trajectories-four-algorithms}
\end{figure}

\begin{table}[t]
\centering
\caption{Comparison of Average Performance Across Four Algorithms for 4-UAV Collaborative Grassland Restoration in All Scenarios.}
\label{UAV4-performance}

\setlength{\tabcolsep}{1.5pt}
\resizebox{0.45\textwidth}{!}{
\begin{tabular}{
c c
c c c >{\columncolor{gray!15}}c}
\toprule

\multirow{2}{*}{\makecell{\textbf{Areas Pending} \\ \textbf{Restoration}}} &
\multirow{2}{*}{\textbf{Metric}} &
\multicolumn{4}{c}{\textbf{Algorithms}} \\
\cmidrule(lr){3-6}
& & \textbf{CAMP} & \textbf{CHAPBILM} & \textbf{MAPDP} & \textbf{KC-BFPRL} \\
\midrule

\multirow{4}{*}{60}
& Obj.  & 675.60 & 798.51 & 862.34 & \textbf{945.80} \\
& Areas    & 612 & 742 & 789 & \textbf{863} \\
& Time (s)         & \textbf{11.21} & 892.49 & 45.79 & 15.81 \\
& Gap (\%)         & 28.59 & 15.57 & 8.83 & \textbf{0.00} \\
\midrule

\multirow{4}{*}{80}
& Obj.  & 878.00 & 1065.42 & 1125.51 & \textbf{1248.27} \\
& Areas    & 798 & 968 & 1023 & \textbf{1134} \\
& Time (s)         & 53.24 & 1287.65 & 62.44 & \textbf{16.85} \\
& Gap (\%)         & 29.65 & 14.64 & 9.83 & \textbf{0.00} \\
\midrule

\multirow{4}{*}{100}
& Obj.  & 1048.18 & 1306.40 & \textbf{1628.46} & 1585.39 \\
& Areas    & 952 & 1187 & \textbf{1478} & 1442 \\
& Time (s)         & 67.82 & 1654.37 & 78.98 & \textbf{21.55} \\
& Gap (\%)         & 35.65 & 19.77 & \textbf{0.00} & 2.64 \\
\midrule

\multirow{4}{*}{120}
& Obj.  & 1272.43 & 1566.26 & 1688.11 & \textbf{1908.76} \\
& Areas    & 1156 & 1423 & 1534 & \textbf{1735} \\
& Time (s)         & \textbf{25.89} & 2021.71 & 94.59 & 28.90 \\
& Gap (\%)         & 33.33 & 17.95 & 11.55 & \textbf{0.00} \\
\midrule

\multirow{4}{*}{140}
& Obj.  & 1526.65 & 1858.53 & 2006.35 & \textbf{2296.12} \\
& Areas    & 1387 & 1689 & 1823 & \textbf{2087} \\
& Time (s)         & 98.52 & 2456.27 & 111.81 & \textbf{31.22} \\
& Gap (\%)         & 33.52 & 19.05 & 12.63 & \textbf{0.00} \\
\midrule

\multirow{4}{*}{160}
& Obj.  & 1798.08 & 2186.17 & \textbf{2756.82} & 2698.50 \\
& Areas    & 1634 & 1987 & \textbf{2504} & 2453 \\
& Time (s)         & 114.22 & 2874.56 & 128.93 & \textbf{36.70} \\
& Gap (\%)         & 34.75 & 20.69 & \textbf{0.00} & 2.12 \\
\bottomrule
\end{tabular}
}
\end{table}

\begin{table}[t]
\centering
\caption{Comparison of Average Performance Across Four Algorithms for 6-UAV Collaborative Grassland Restoration in All Scenarios.}
\label{UAV6-performance}

\setlength{\tabcolsep}{1.5pt}
\resizebox{0.45\textwidth}{!}{
\begin{tabular}{
c c
c c c >{\columncolor{gray!15}}c}
\toprule

\multirow{2}{*}{\makecell{\textbf{Areas Pending} \\ \textbf{Restoration }}} &
\multirow{2}{*}{\textbf{Metric}} &
\multicolumn{4}{c}{\textbf{Algorithms}} \\
\cmidrule(lr){3-6}
& & \textbf{CAMP} & \textbf{CHAPBILM} & \textbf{MAPDP} & \textbf{KC-BFPRL} \\
\midrule

\multirow{4}{*}{60}
& Obj.  & 1010.59 & 1198.18 & 1306.18 & \textbf{1425.49} \\
& Areas    & 918  & 1089 & 1187 & \textbf{1295} \\
& Time (s)         & \textbf{16.82}   & 1340.78 & 68.98   & 22.74   \\
& Gap (\%)         & 29.10  & 15.95  & 8.37   & \textbf{0.00}   \\
\midrule

\multirow{4}{*}{80}
& Obj.  & 1317.09 & 1528.79 & \textbf{1898.60} & 1834.28 \\
& Areas    & 1197 & 1389 & \textbf{1726} & 1666 \\
& Time (s)         & 76.57   & 1789.31 & 89.21   & \textbf{24.74}   \\
& Gap (\%)         & 30.63  & 19.48  & \textbf{0.00}   & 3.39   \\
\midrule

\multirow{4}{*}{100}
& Obj.  & 1572.18 & 1907.65 & 2116.23 & \textbf{2348.22} \\
& Areas    & 1429 & 1734 & 1923 & \textbf{2134} \\
& Time (s)         & 95.88   & 2287.49 & 112.81  & \textbf{31.24}   \\
& Gap (\%)         & 33.05  & 18.77  & 9.88   & \textbf{0.00}   \\
\midrule

\multirow{4}{*}{120}
& Obj.  & 1908.63 & 2298.59 & 2528.63 & \textbf{2858.27} \\
& Areas    & 1735 & 2089 & 2298 & \textbf{2598} \\
& Time (s)         & 38.96   & 2798.62 & 136.71  & 41.54   \\
& Gap (\%)         & 33.22  & 19.59  & 11.53  & \textbf{0.00}   \\
\midrule

\multirow{4}{*}{140}
& Obj.  & 2292.66 & 2702.57 & 3007.60 & \textbf{3398.74} \\
& Areas    & 2084 & 2456 & 2734 & \textbf{3089} \\
& Time (s)         & 135.88  & 3356.91 & 161.33  & \textbf{45.11}   \\
& Gap (\%)         & 32.56  & 20.48  & 11.51  & \textbf{0.00}   \\
\midrule

\multirow{4}{*}{160}
& Obj.  & 2695.76 & 3154.12 & 3518.65 & \textbf{3998.35} \\
& Areas    & 2450 & 2867 & 3198 & \textbf{3634} \\
& Time (s)         & \textbf{54.21}   & 3897.26 & 186.84  & 56.82   \\
& Gap (\%)         & 32.57  & 21.12  & 12.00  & \textbf{0.00}   \\
\bottomrule

\end{tabular}
}
\end{table}

\begin{table}[t]
\centering
\caption{Comparison of Average Performance Across Four Algorithms for 8-UAV Collaborative Grassland Restoration in All Scenarios.}
\label{UAV8-performance}

\setlength{\tabcolsep}{1.5pt}

\resizebox{0.45\textwidth}{!}{
\begin{tabular}{
c c
c c c >{\columncolor{gray!15}}c}
\toprule

\multirow{2}{*}{\makecell{\textbf{Areas Pending} \\ \textbf{Restoration}}} &
\multirow{2}{*}{\textbf{Metric}} &
\multicolumn{4}{c}{\textbf{Algorithms}} \\
\cmidrule(lr){3-6}
& & \textbf{CAMP} & \textbf{CHAPBILM} & \textbf{MAPDP} & \textbf{KC-BFPRL} \\
\midrule

\multirow{4}{*}{60}
& Obj.  & 1347.36 & 1428.41 & \textbf{1789.27} & 1718.95 \\
& Areas    & 1224 & 1298 & \textbf{1625} & 1562 \\
& Time (s)         & 69.77   & 1643.51 & 82.11   & \textbf{22.64}   \\
& Gap (\%)         & 24.69  & 20.16  & \textbf{0.00}   & 3.93   \\
\midrule

\multirow{4}{*}{80}
& Obj.  & 1756.63 & 1868.25 & 2076.01 & \textbf{2298.61} \\
& Areas    & 1596 & 1698 & 1887 & \textbf{2089} \\
& Time (s)         & \textbf{29.87}   & 2156.76 & 107.32  & 32.15   \\
& Gap (\%)         & 23.58  & 18.71  & 9.68   & \textbf{0.00}   \\
\midrule

\multirow{4}{*}{100}
& Obj.  & 2097.20 & 2298.57 & 2604.63 & \textbf{2898.23} \\
& Areas    & 1906 & 2089 & 2367 & \textbf{2634} \\
& Time (s)         & 114.52  & 2734.87 & 134.90  & \textbf{37.52}   \\
& Gap (\%)         & 27.64  & 20.69  & 10.13  & \textbf{0.00}   \\
\midrule

\multirow{4}{*}{120}
& Obj.  & 2544.18 & 2776.27 & 3154.13 & \textbf{3518.46} \\
& Areas    & 2312 & 2523 & 2867 & \textbf{3198} \\
& Time (s)         & \textbf{42.19}   & 3287.69 & 162.83  & 48.71   \\
& Gap (\%)         & 27.69  & 21.09  & 10.35  & \textbf{0.00}   \\
\midrule

\multirow{4}{*}{140}
& Obj.  & 2990.23 & 3264.11 & 3726.34 & \textbf{4178.61} \\
& Areas    & 2718 & 2967 & 3387 & \textbf{3798} \\
& Time (s)         & 163.24  & 3865.97 & 192.11  & \textbf{53.48}   \\
& Gap (\%)         & 28.45  & 21.88  & 10.83  & \textbf{0.00}   \\
\midrule

\multirow{4}{*}{160}
& Obj.  & 3547.14 & 3835.99 & 4386.42 & \textbf{4914.08} \\
& Areas    & 3224 & 3487 & 3987 & \textbf{4467} \\
& Time (s)         & \textbf{65.19}   & 4453.23 & 222.46  & 68.54   \\
& Gap (\%)         & 27.82  & 21.93  & 10.74  & \textbf{0.00}   \\
\bottomrule

\end{tabular}
}
\end{table}
We evaluate our method against CHAPBILM~\cite{JIAO2024108084}, MAPDP~\cite{zong2022mapdp}, and CAMP~\cite{hua2025camp}. The experimental results for fleet sizes of U4, U6, and U8 are presented in Tables \ref{UAV4-performance}, \ref{UAV6-performance}, and \ref{UAV8-performance}, respectively.
The optimality gap is defined as the normalized difference between the average objective value of a given method (Obj.) and the best objective value ($\text{Obj}_{\text{best}}$) across all methods:
$\text{Gap} = \frac{Obj_{\text{best}} - Obj}{Obj_{\text{best}}} \times 100\%$.

Tables~\ref{UAV4-performance} to \ref{UAV8-performance} show that KC-BFPRL consistently outperforms baselines across all fleet sizes and scales, particularly in large-scale scenarios involving 120 to 160 regions, with the minor exception of U4-R160. In the complex U8-R160 instance, it surpasses MAPDP and CAMP by $12.03\%$ in objective value and approximately $38.55\%$ in total restored areas, respectively. While MAPDP’s optimality gap widens to $10.74\%$ as complexity increases, KC-BFPRL maintains a  $0.00\%$ gap across all 8-UAV instances. Furthermore, it is nearly three times faster than MAPDP, effectively mitigating the ``curse of dimensionality". Its robustness is highlighted by maintaining a mean optimality gap under $2\%$ as the fleet scales.

\begin{figure}[htbp]
    \centering
    \subfigure[Energy efficiency]{\includegraphics[width=1.2in]{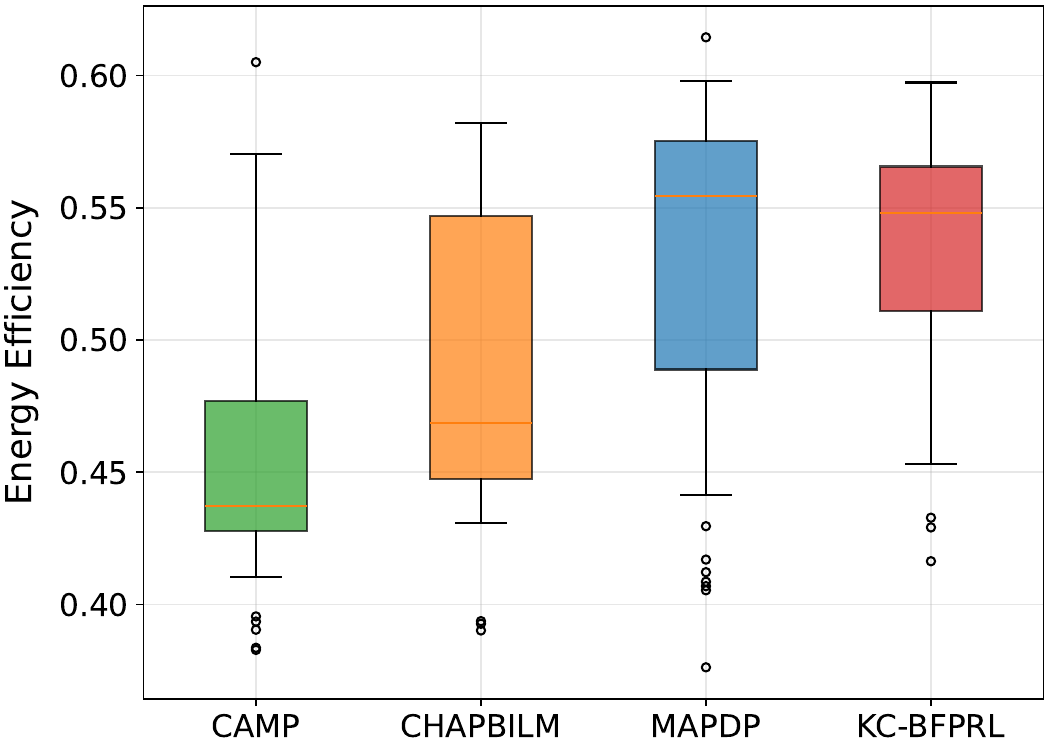}}
    \subfigure[Number of restoration areas]{\includegraphics[width=1.2in]{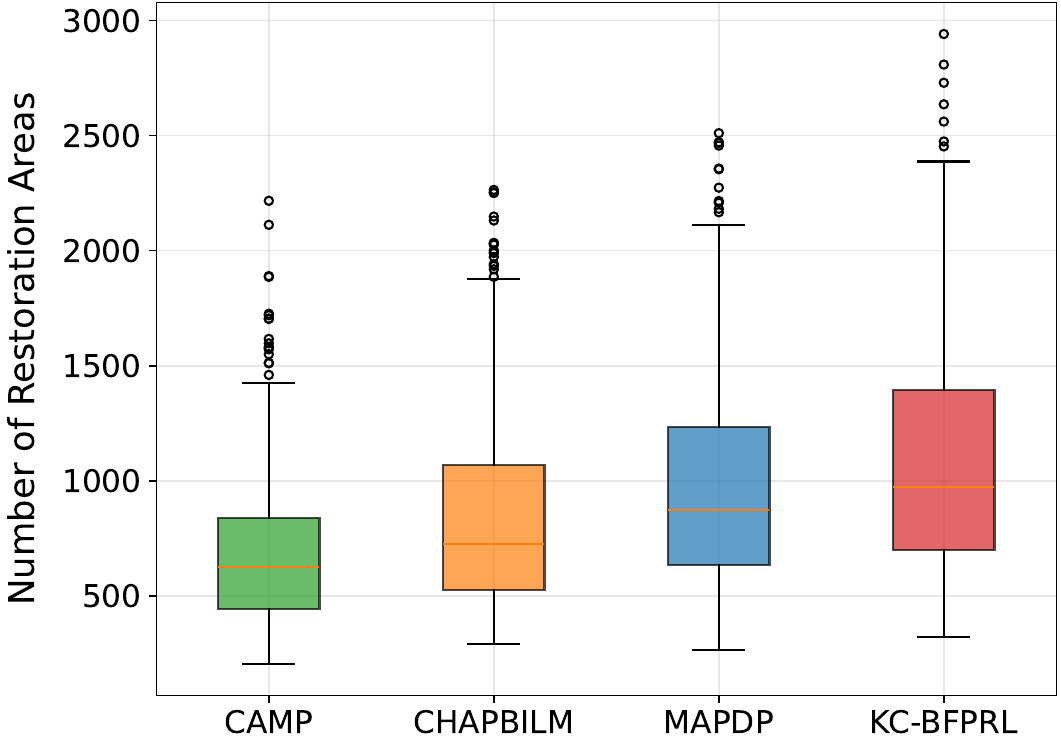}}
    \subfigure[Trajectory length]{\includegraphics[width=1.2in]{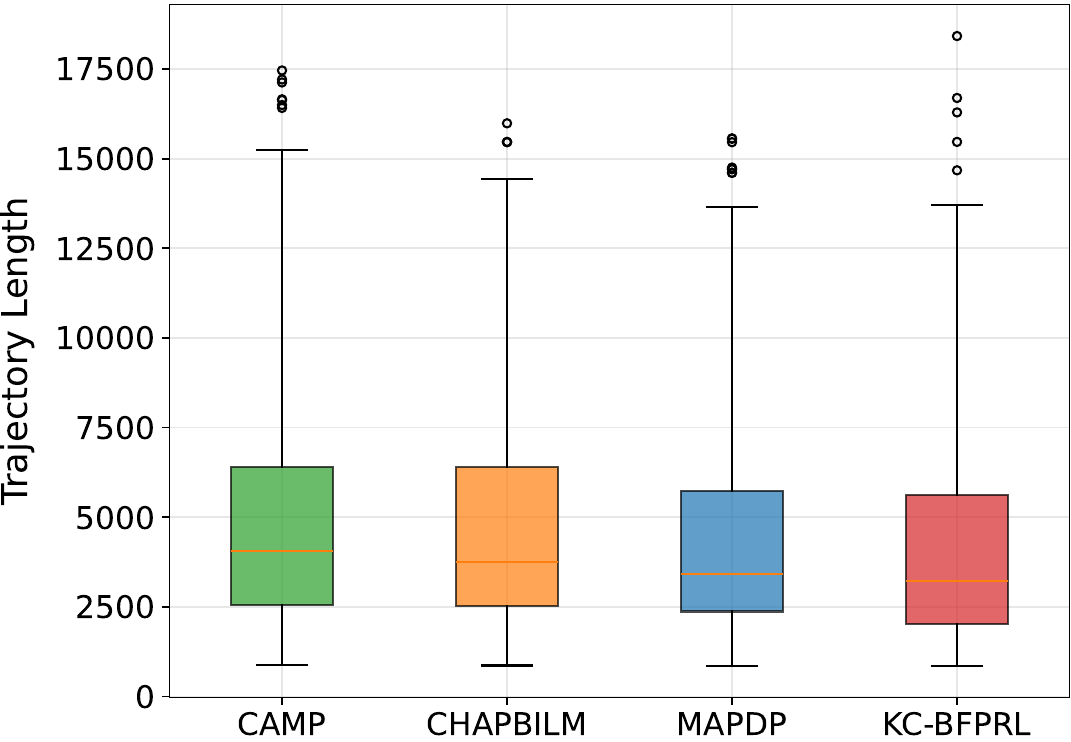}}
    \subfigure[Objective function value]{\includegraphics[width=1.2in]{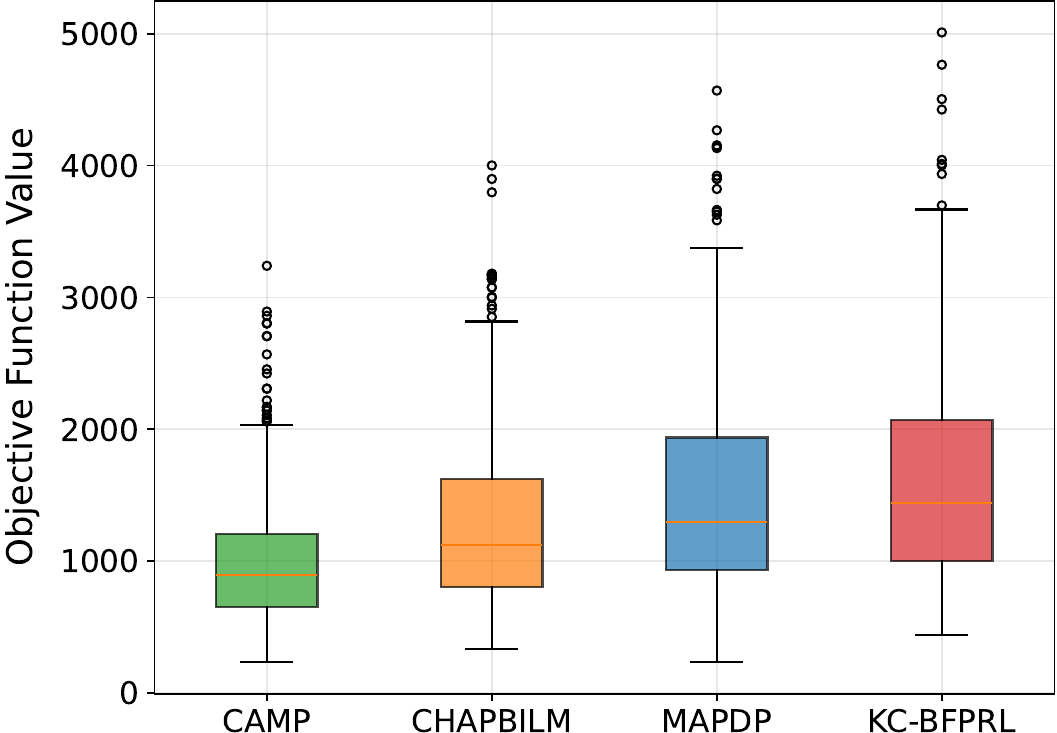}}
\caption{Comparative performance of the four algorithms on four key metrics.} \label{Boxplot}
\end{figure}

Statistical results in presented Fig. \ref{Boxplot} and Table \ref{average-performance-comparison} of Appendix \ref{extended-experimental} demonstrate  the superiority of KC-BFPRL across four key metrics. It achieves the highest average restoration areas of 1109.73, outperforming MAPDP, CHAPBILM, and CAMP by $11.60\%$, $28.09\%$, and $58.52\%$, respectively. It also yields the most efficient flight paths, with an average length of 4068.40, representing distance reductions of $3.69\%$, $12.83\%$, and $13.56\%$ compared to the baselines. Moreover, KC-BFPRL exhibits higher median and minimum values along with lower standard deviations in energy efficiency, resulting in a mean value of 0.538 that underscores its exceptional stability. Finally, KC-BFPRL surpasses its closest competitor, MAPDP, by $8.90\%$ in overall objective value, validating its practical engineering value for large-scale, complex ecological restoration tasks.

\subsubsection{Generalization and Robustness Study}
\begin{figure}[htb]
    \centering
    \subfigure[U4-R120]{\includegraphics[width=1.0in]{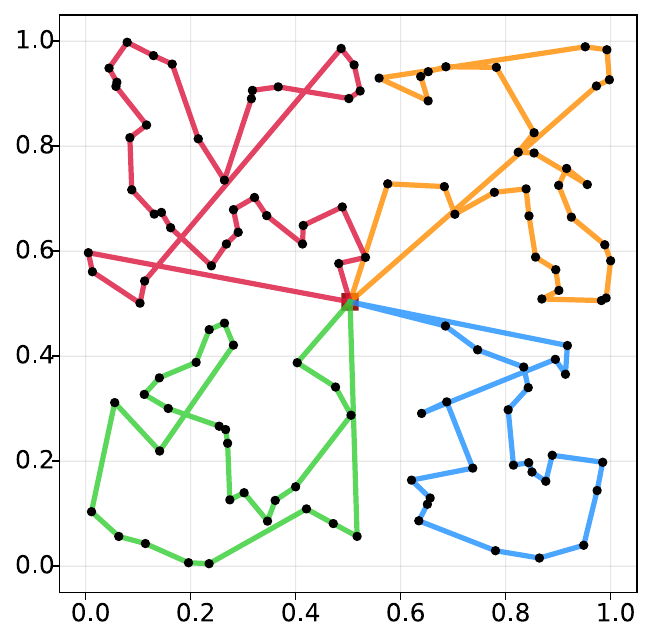}}
    \subfigure[U6-R120]{\includegraphics[width=1.0in]{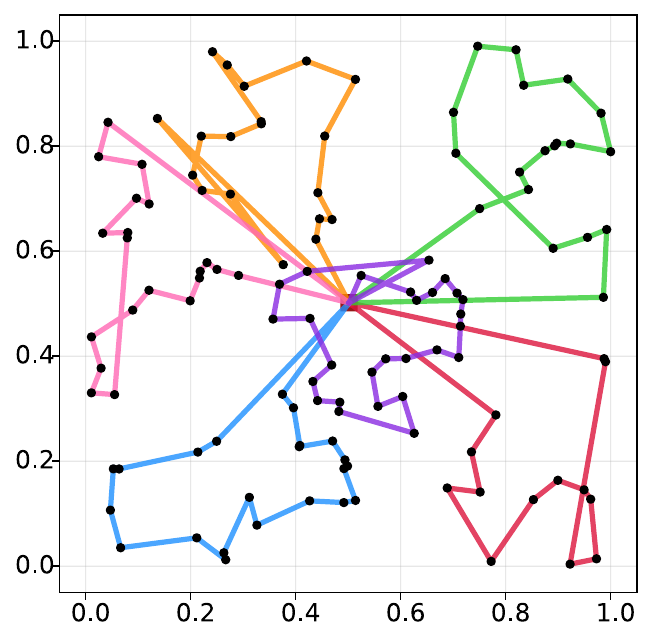}}
    \subfigure[U8-R120]{\includegraphics[width=1.0in]{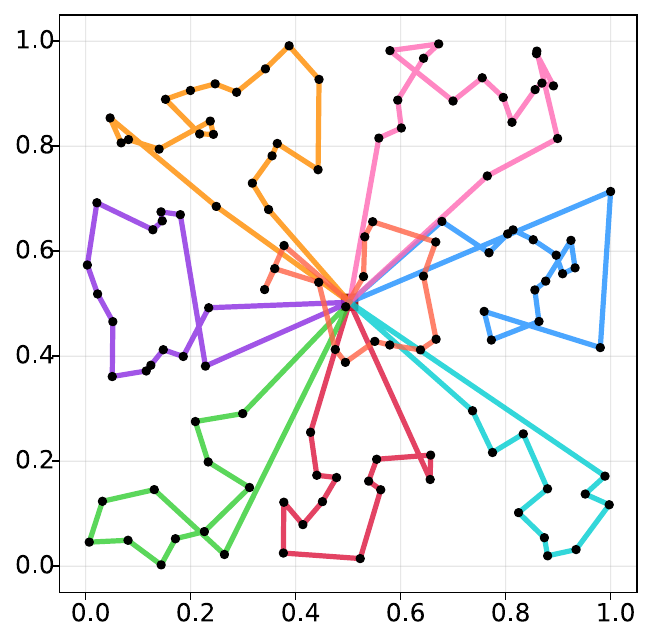}}
    \subfigure[U4-R140]{\includegraphics[width=1.0in]{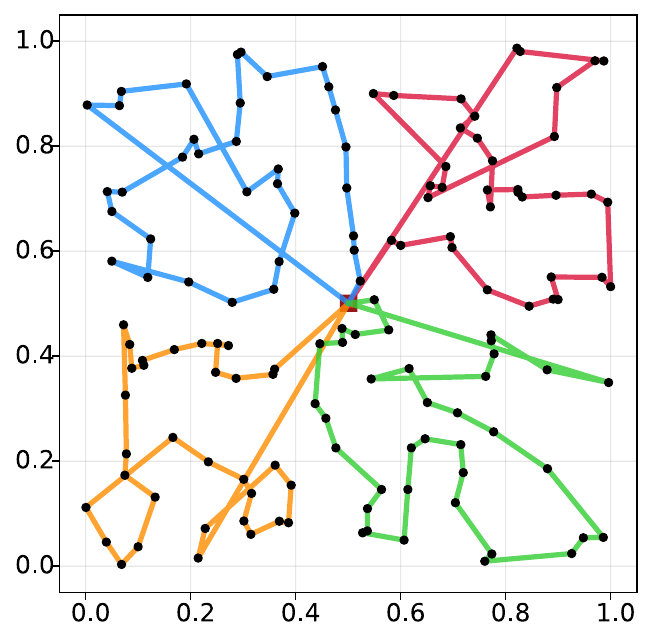}}
    \subfigure[U6-R140]{\includegraphics[width=1.0in]{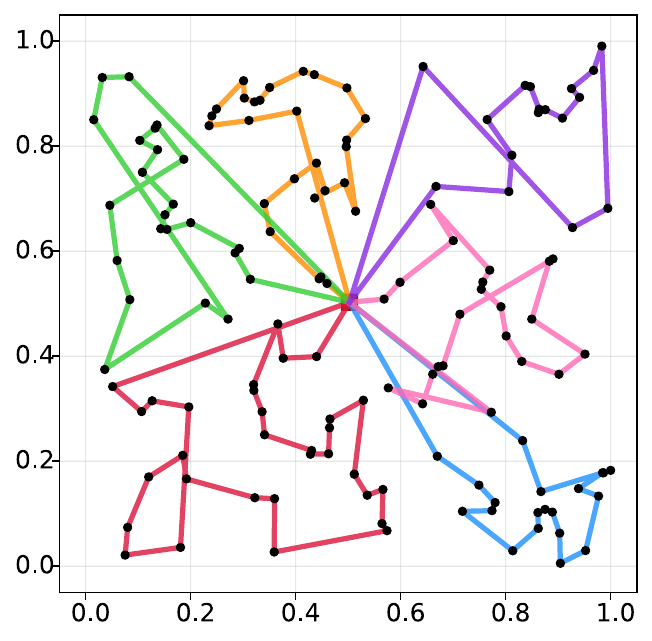}}
    \subfigure[U8-R140]{\includegraphics[width=1.0in]{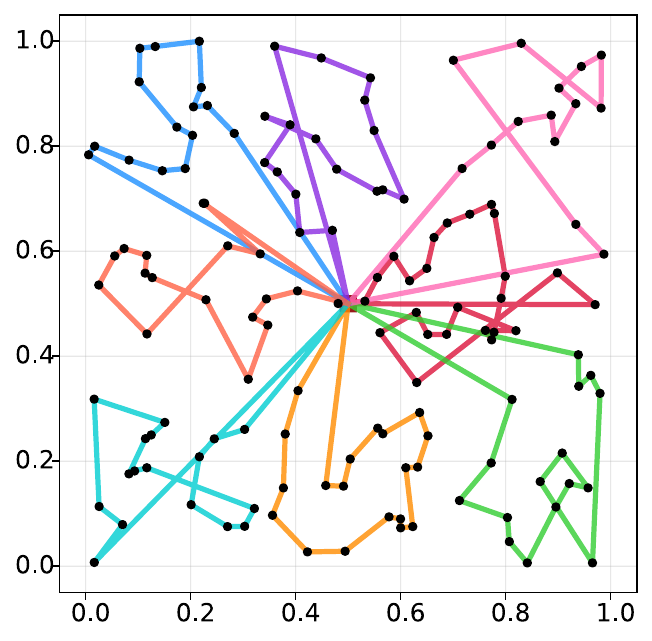}}
    \subfigure[U4-R160]{\includegraphics[width=1.0in]{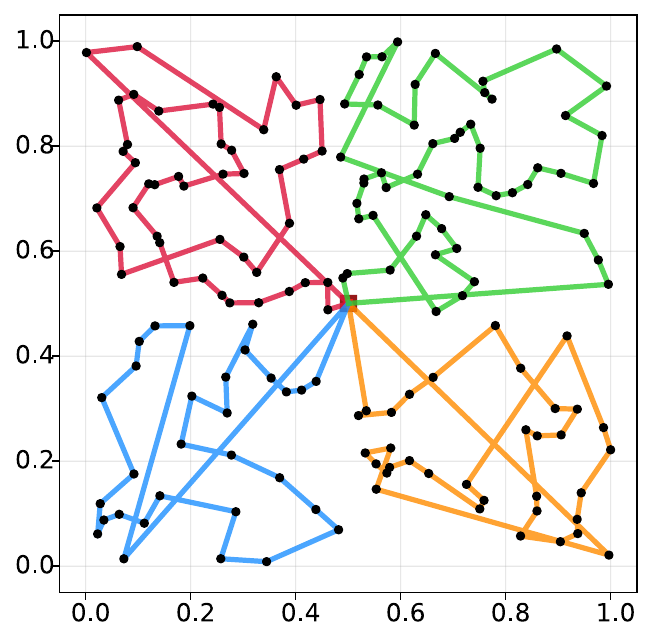}}
    \subfigure[U6-R160]{\includegraphics[width=1.0in]{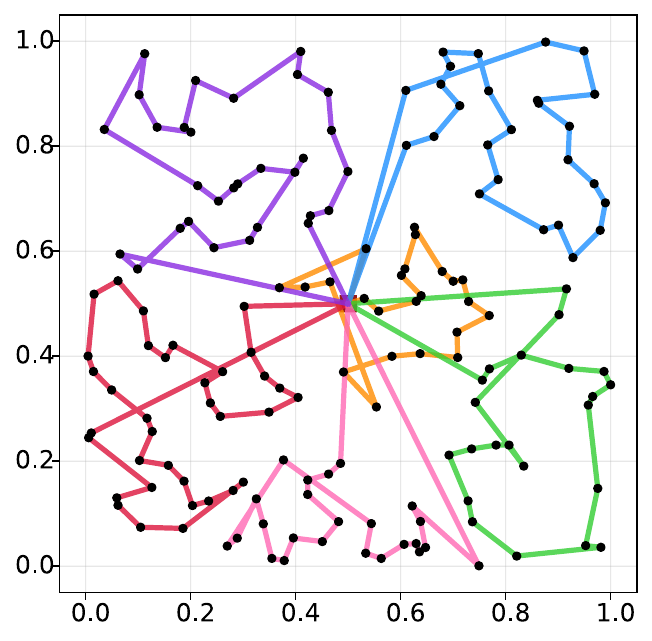}}
    \subfigure[U8-R160]{\includegraphics[width=1.0in]{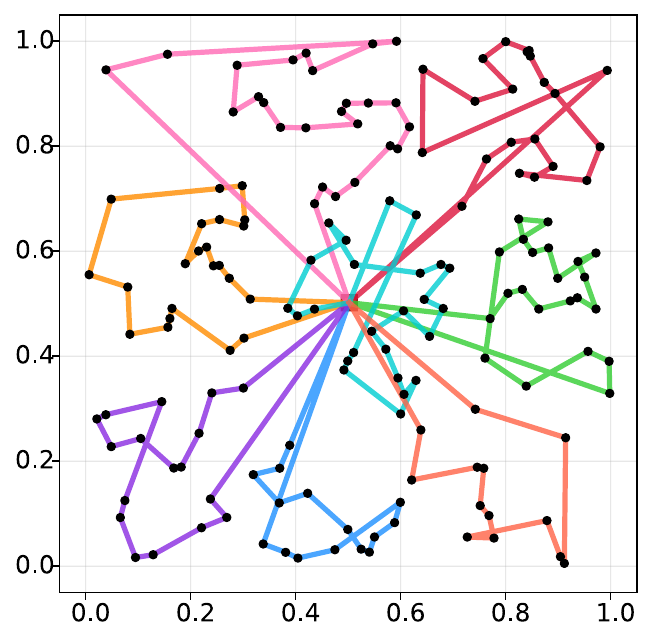}}
\caption{Visualization of optimal KC-BFPRL trajectories for 4, 6, and
8 UAVs across three medium-to-large-scale scenarios.} \label{Optimal-KC-BFPRL-trajectories-medium-to-large-scale}
\end{figure}

We evaluate the model’s generalization by testing a fixed policy across varying problem sizes without fine-tuning. Fig. \ref{Optimal-KC-BFPRL-trajectories-medium-to-large-scale} shows that KC-BFPRL maintains high-quality solutions even in mismatched settings with minimal performance loss. Notably, the pre-trained model often outperforms baselines specifically trained for those scenarios. These results confirm the versatility of our knowledge-guided framework and its potential for broad applicability in complex multi-UAV challenges.

\subsubsection{Discussion}
Based on the comparative performance analysis, we delineate the specific applicability of each algorithm to different operational contexts. KC-BFPRL balances optimality and efficiency, making it ideal for large-scale, complex missions involving over 100 restoration regions that require real-time decision-making. In contrast, MAPDP is suitable for precision-critical tasks where latency is acceptable, while CHAPBILM is best for smaller, training-free offline scenarios with fewer than 80 regions.  KC-BFPRL’s success lies in its knowledge-guided paradigm, which prunes the search space using ecological priority, heuristic logic, and hierarchical coordination. This ``warm-start" avoids learning from scratch and mitigates the ``curse of dimensionality.” Furthermore, by utilizing transparent heuristics, the framework enhances interpretability and provides a systematic guide for selecting algorithms based on mission scale and resource constraints.

\section{Conclusion} \label{Conclusion}
This paper addressed the critical challenge of multi-UAV collaborative grassland restoration by formulating the RAMP and proposing the KC-BFPRL framework. By integrating domain expertise, specifically ecological priority and heuristic scheduling logic, with the adaptive capability of DRL, the proposed approach effectively bridges the gap between centralized coordination and decentralized execution. Extensive experiments demonstrates that KC-BFPRL significantly outperforms state-of-the-art heuristic and learning-based baselines, successfully enabling real-time, high-precision decision-making. Essentially, the knowledge-guided paradigm ensures robust scalability and interpretability, maintaining near-optimal performance even as task complexity increases. Future work will explore problem-decomposition for lower latency and conduct real-world trials under dynamic environmental uncertainties to advance autonomous ecological engineering.

\appendices
\section{Hyperparameter Configuration for Multi-Agent Reinforcement Learning Methods}\label{Hyperparameter}
All multi-agent reinforcement learning (MARL) baselines were implemented using the source code and hyperparameters provided in their original publications. To ensure an equitable comparison, each baseline model was trained for 10000 epochs, consistent with the training protocol established for KC-BFPRL. Because the native architectures of MAPDP and CAMP cannot be directly applied to RAMP, we adapted their input representations and constraint-handling mechanisms to align with our mathematical formulation. Specifically, we explicitly integrated the critical constraints governing spatially heterogeneous degradation levels and dynamic UAV energy limitations into their decision-making processes. The specific parameter settings are shown in the table~\ref{parameter-settings}.
\begin{table}[htbp]
\centering
\caption{Transformer model training parameters} \label{parameter-settings}
\setlength{\tabcolsep}{4pt}
\begin{tabular}{lll}
\toprule
Parameter & Description & Value \\
\midrule
$n_\text{layers}$ & number of Transformer layers & 3 \\
$d_\text{in}$ & input dimension & 4 ($x$, $y$, $l$, $R$) \\
$d_e$ & embedding dimension & 128 \\
$C$ & attention scaling factor & 10 \\
$N_\text{epoch}$ & number of epochs & 100 \\
$B$ & batch size & 256 \\
$\alpha$ & learning rate (Adam) & $3 \times 10^{-4}$ \\
$g_\text{max}$ & gradient clipping threshold & 0.5 \\
$D_\text{val}$ & validation set size & 256 \\
$T_\text{epochs}$ & number of epochs (training) & 10000 \\
$\alpha_p$ & penalty term weight     & 10.0 \\
$\alpha_r$ & recovery term weight    & 1.0 \\
\bottomrule
\end{tabular}
\end{table}
\section{Extended Experimental Results} \label{extended-experimental}
\begin{figure}[tbp]
\begin{center}
$\begin{array}{l}
\includegraphics[width=3.0in]{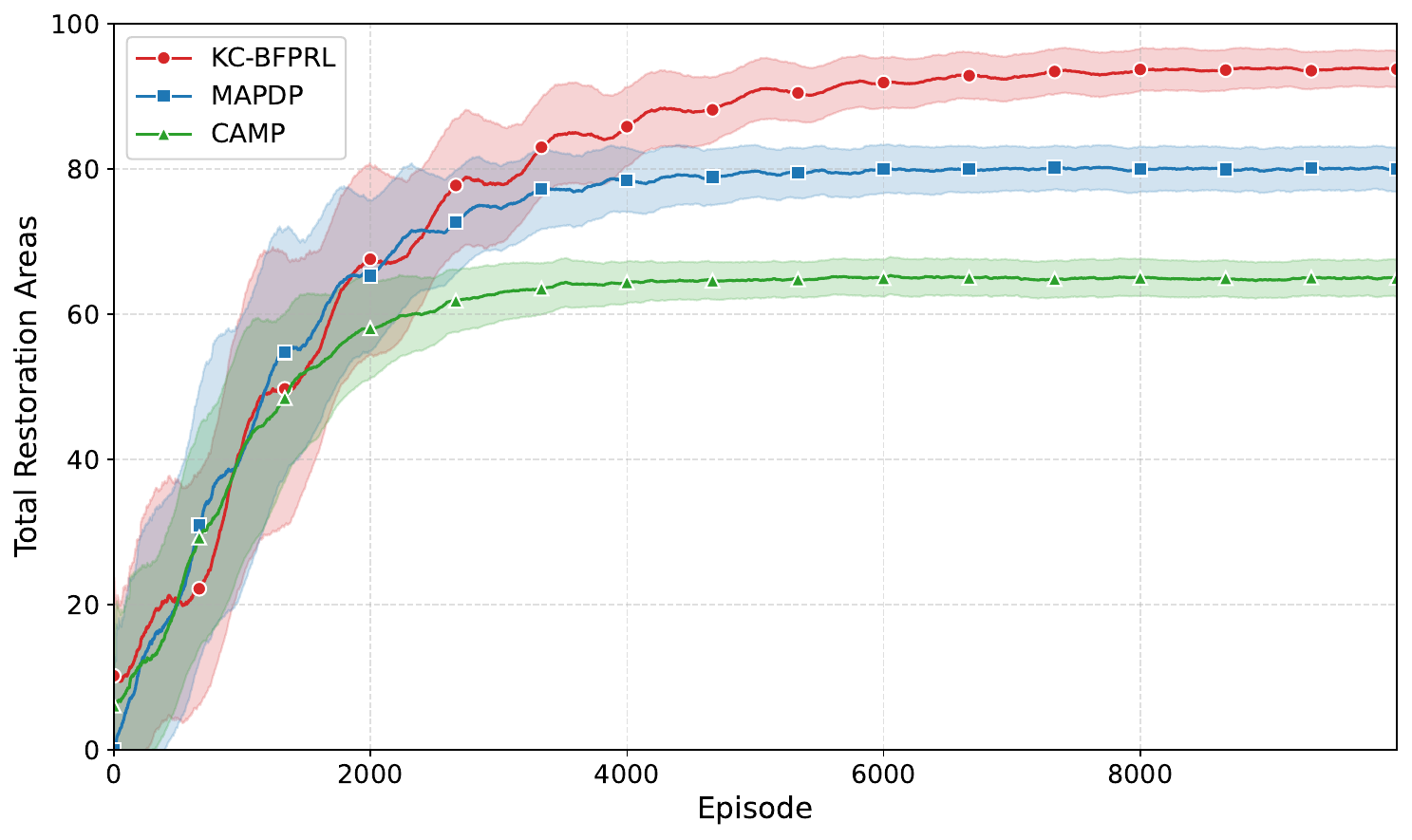}\\
 \end{array}$
\end{center}
\vspace{-0.15in}
\caption{Evolution of total restoration areas across training epochs for the three learning-based algorithms in scenario U4-R60.} \label{Total-Area-Training-Episodes}
\end{figure}

Fig. \ref{Total-Area-Training-Episodes} illustrates the accumulation of total restored areas across training epochs, comparing the learning curves of the three learning-based approaches under the U4-R60 configuration.

\begin{figure}[tb]
    \centering
    \subfigure[U4-R60]{\includegraphics[width=1.1in]{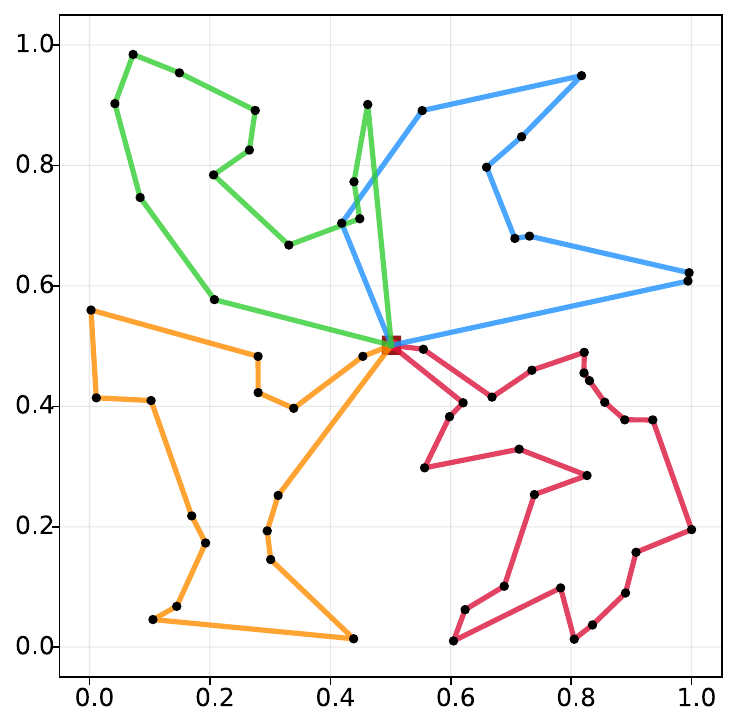}}
    \subfigure[U6-R60]{\includegraphics[width=1.1in]{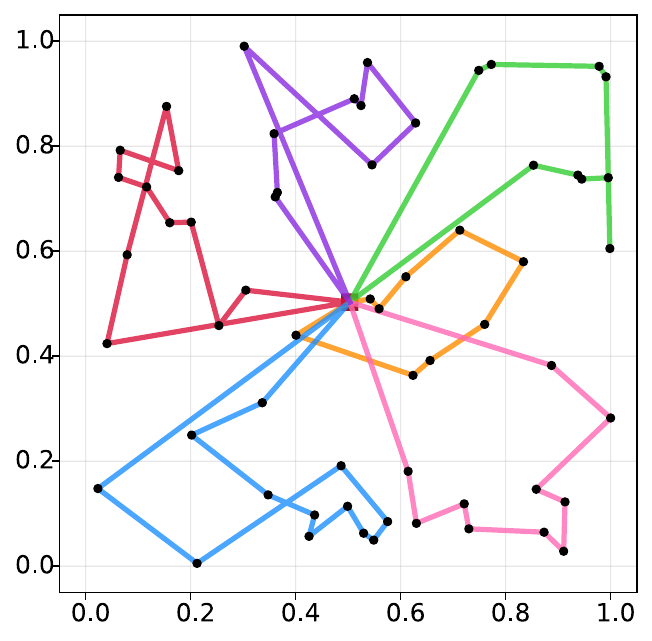}}
    \subfigure[U8-R60]{\includegraphics[width=1.1in]{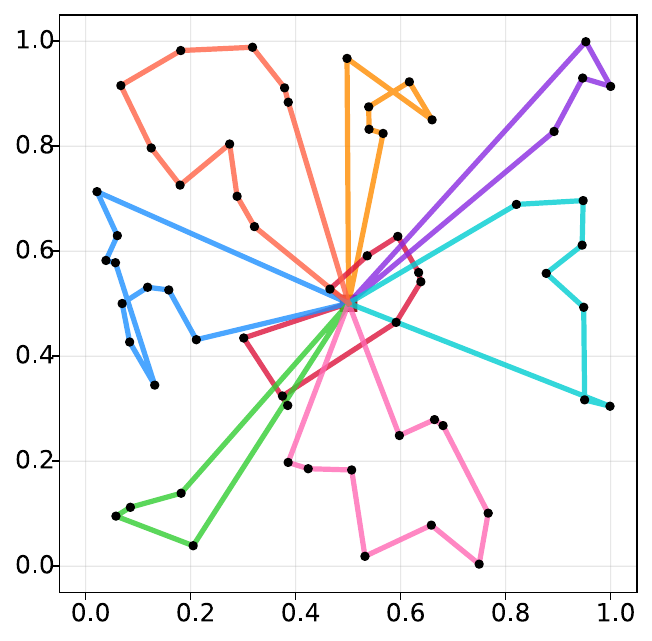}}
    \subfigure[U4-R80]{\includegraphics[width=1.1in]{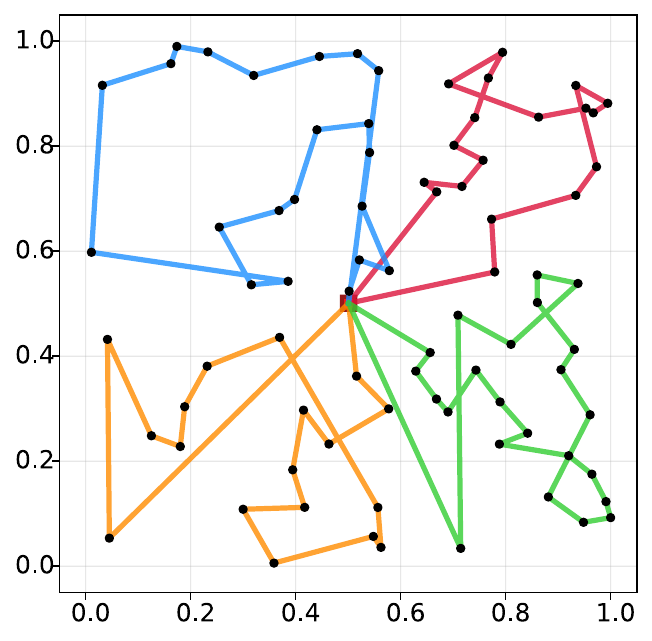}}
    \subfigure[U6-R80]{\includegraphics[width=1.1in]{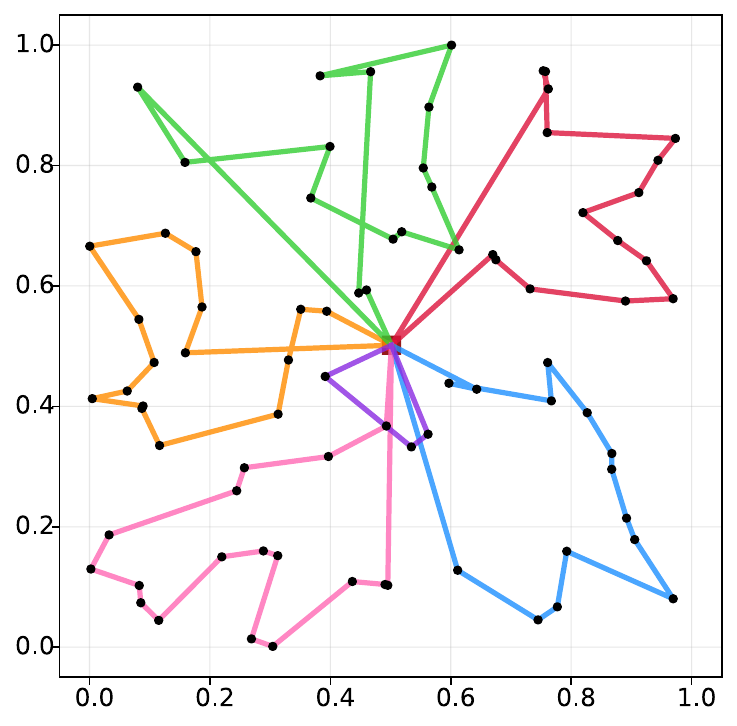}}
    \subfigure[U8-R80]{\includegraphics[width=1.1in]{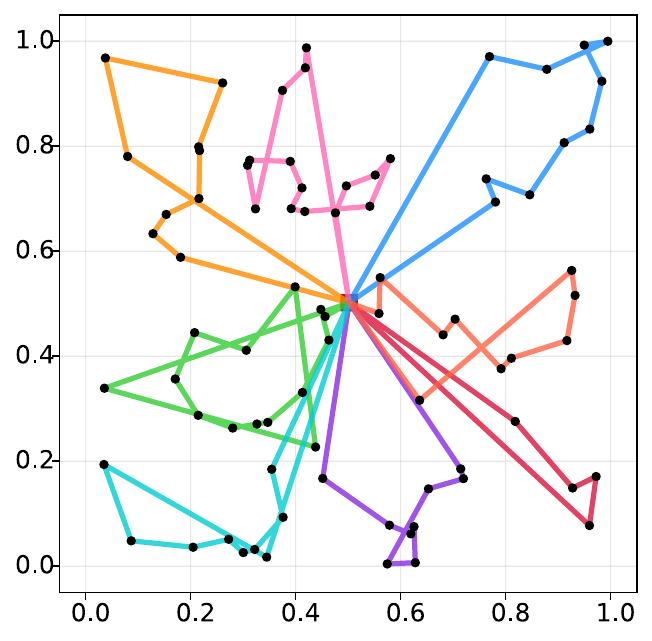}}
    \subfigure[U4-A100]{\includegraphics[width=1.1in]{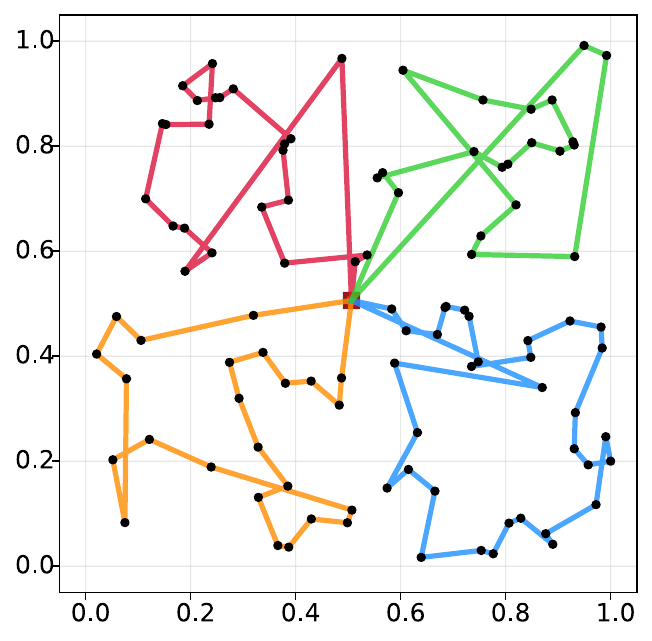}}
    \subfigure[U6-R100]{\includegraphics[width=1.1in]{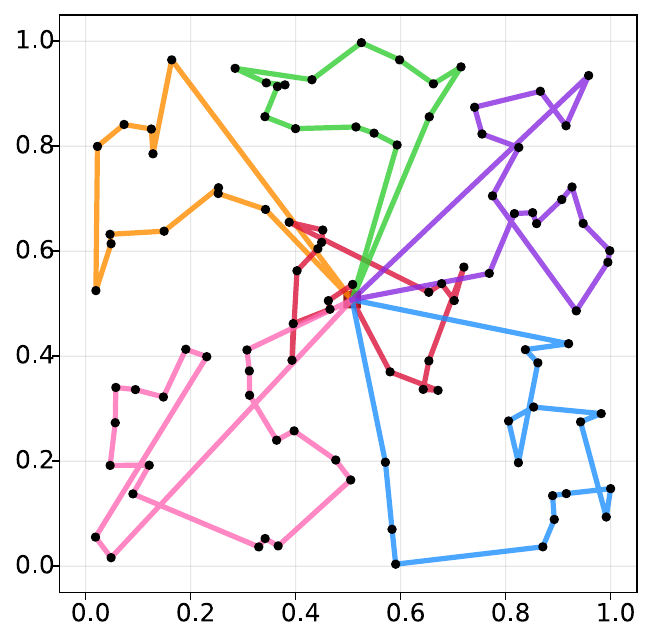}}
    \subfigure[U8-R100]{\includegraphics[width=1.1in]{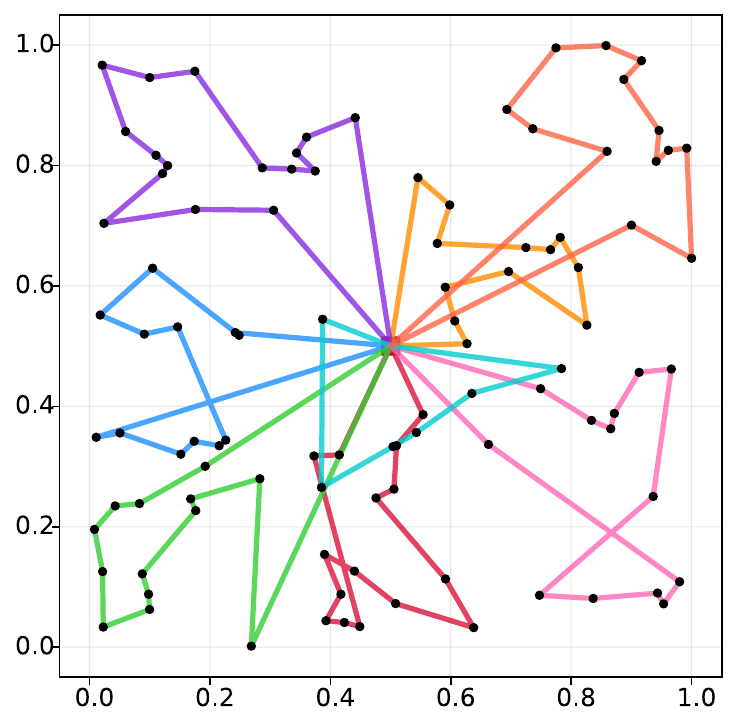}}
\caption{Visualization of optimal KC-BFPRL trajectories for 4, 6, and 8 UAVs across three small-to-medium-scale scenarios.} \label{Optimal-KC-BFPRL-trajectories-small-to-medium-scale}
\end{figure}

Fig. \ref{Optimal-KC-BFPRL-trajectories-small-to-medium-scale} visualizes the optimal trajectories generated by the proposed KC-BFPRL framework. The rows correspond to varying fleet sizes (4, 6, and 8 UAVs), while the columns denote three distinct medium-to-large-scale scenarios. This layout provides a clear qualitative assessment of the algorithm's path-planning performance across different mission scales and spatial constraints.

Table~\ref{average-performance-comparison} presents a comprehensive comparison of the average performance metrics achieved by the four algorithms. The reported values are averaged across all test instances, providing a consolidated evaluation of each algorithm's relative effectiveness across the primary criteria.
\begin{table}[t]
\centering
\caption{Comparison of average performance metrics across the four algorithms.}
\label{average-performance-comparison}
\setlength{\tabcolsep}{2pt}
\resizebox{0.48\textwidth}{!}{
\begin{tabular}{llcccccc}
\toprule
\textbf{Metric} & \textbf{Algorithm} & \textbf{Mean} & \textbf{Std} & \textbf{Median} & \textbf{Min} & \textbf{Max} & \textbf{Q75} \\
\midrule\arrayrulecolor{black}

\multirow{4}{*}{\makecell{\cellcolor{white}Restoration\\Areas}}
& CAMP     & 700 & \textbf{359} & 625  & 203 & 2216 & 838 \\
& CHAPBILM & 866 & 449 & 725  & 292 & 2263 & 1067 \\
& MAPDP    & 994 & 487 & 872  & 262 & 2510 & 1232 \\
\rowcolor{gray!15}\cellcolor{white}& KC-BFPRL & \textbf{1109} & 544 & \textbf{971} & \textbf{322} & \textbf{2941} & \textbf{1395} \\
\midrule\arrayrulecolor{black}

\multirow{4}{*}{\makecell{\cellcolor{white}Energy\\Efficiency}}
& CAMP     & 0.457 & 0.043 & 0.437 & 0.411 &0.570 &0.477 \\
& CHAPBILM & 0.494 & 0.052 & 0.469 & 0.431 & 0.582 & 0.547 \\
& MAPDP    & 0.537 & 0.047 & \textbf{0.555} & 0.441 & \textbf{0.598} & \textbf{0.575} \\
\rowcolor{gray!15}\cellcolor{white}& KC-BFPRL & \textbf{0.538} & \textbf{0.035} & 0.548 & \textbf{0.453} & 0.597 & 0.566 \\
\midrule\arrayrulecolor{black}

\multirow{4}{*}{\makecell{\cellcolor{white}Trajectory\\Length}}
& CAMP     & 4706.47 & 2677.40 & 4052.19 & \textbf{853.68} & 13725.89 & 6173.27 \\
& CHAPBILM & 4667.37 & 2716.91 & 3669.33 & 859.86 & 13507.69 & 6631.64 \\
& MAPDP    & 4224.46 & 2561.48 & 3433.10 & 949.60 & 13821.66 & 5639.78 \\
\rowcolor{gray!15}\cellcolor{white}& KC-BFPRL & \textbf{4068.40} & \textbf{2379.57} & \textbf{3419.49} & 918.53 &\textbf{12655.74}&\textbf{5509.92} \\
\midrule\arrayrulecolor{black}

\multirow{4}{*}{\makecell{\cellcolor{white}Objective\\Value}}
& CAMP     & 1024.52 & \textbf{543.90} & 893.03 & 233.07 & 3239.36 & 1204.10 \\
& CHAPBILM & 1313.71 & 712.87 & 1121.74 & 334.85 & 4000.82 & 1620.81 \\
& MAPDP    & 1515.66 & 829.19 & 1296.49 & 230.85 & 4568.51 & 1936.94 \\
\rowcolor{gray!15}\cellcolor{white}& KC-BFPRL & \textbf{1650.54} & 865.87 & \textbf{1437.29} & \textbf{442.23} & \textbf{5009.81} & \textbf{2073.11} \\
\bottomrule
\end{tabular}
}
\end{table}


%

%

\section*{Acknowledgment}
This work was supported in part by the National Natural Science Foundation of China under Grant 62233003 and 62272210, and in part by the Natural Science Foundation of Gansu Province under Grant 24JRRA430. We thank Mr. B. Liu for his assistance and the anonymous reviewers for their constructive feedback which improved this manuscript.

\ifCLASSOPTIONcaptionsoff
  \newpage
\fi



%
%
%

\bibliographystyle{IEEEtran}

\bibliography{Reference}

%

%
%
%




\end{document}